\documentclass[letterpaper,twocolumn,prl,aps,superscriptaddress,amsmath,amssymb,floatfix]{revtex4-1}
\usepackage{mathptmx}
\usepackage[latin9]{inputenc}
\usepackage{color}
\usepackage{amsmath}
\usepackage{amssymb}
\usepackage{graphicx}
\usepackage{esint}
\usepackage[unicode=true,
 bookmarks=true,bookmarksnumbered=false,bookmarksopen=false,
 breaklinks=false,pdfborder={0 0 1},backref=false,colorlinks=true]
 {hyperref}
\hypersetup{
 linkcolor=magenta,urlcolor=blue,citecolor=blue,pdfstartview={FitH}}

\makeatletter

\usepackage{textcomp}
\usepackage{epstopdf}

\usepackage{amsfonts}

\pdfpageheight\paperheight
\pdfpagewidth\paperwidth

\@ifundefined{textcolor}{}{%
 \definecolor{BLACK}{gray}{0}
 \definecolor{WHITE}{gray}{1}
 \definecolor{RED}{rgb}{1,0,0}
 \definecolor{GREEN}{rgb}{0,1,0}
 \definecolor{BLUE}{rgb}{0,0,1}
 \definecolor{CYAN}{cmyk}{1,0,0,0}
 \definecolor{MAGENTA}{cmyk}{0,1,0,0}
 \definecolor{YELLOW}{cmyk}{0,0,1,0}
}
\usepackage{xcolor}\usepackage{soul}
\definecolor{blue}{rgb}{0,0,1}
\definecolor{red}{rgb}{1,0,0}
\definecolor{green}{rgb}{0,1,0}

\usepackage{soul}

\makeatother

\begin{document}

\affiliation{Laboratory of Quantum Information, University of Science and Technology of China, Hefei 230026, China.}
\affiliation{School of Information Science and Technology, ShanghaiTech University, Shanghai 201210, China.}
\affiliation{CAS Center For Excellence in Quantum Information and Quantum Physics,
University of Science and Technology of China, Hefei, Anhui 230026,
China.}
\affiliation{Hefei National Laboratory, University of Science and Technology of China, Hefei 230088, China}

\title{Autonomous frequency locking for zero-offset microcombs}

\author{Fengyan Yang}
\thanks{These authors contributed equally to this work.}
\affiliation{Laboratory of Quantum Information, University of Science and Technology of China, Hefei 230026, China.}
\affiliation{CAS Center For Excellence in Quantum Information and Quantum Physics,
University of Science and Technology of China, Hefei, Anhui 230026,
China.}

\author{Ming Li}
\thanks{These authors contributed equally to this work.}
\affiliation{Laboratory of Quantum Information, University of Science and Technology of China, Hefei 230026, China.}
\affiliation{CAS Center For Excellence in Quantum Information and Quantum Physics,
University of Science and Technology of China, Hefei, Anhui 230026,
China.}
\affiliation{Hefei National Laboratory, University of Science and Technology of China, Hefei 230088, China}

\author{Juanjuan Lu}
\affiliation{School of Information Science and Technology, ShanghaiTech University, Shanghai 201210, China.}

\author{Guang-Can Guo}
\affiliation{Laboratory of Quantum Information, University of Science and Technology of China, Hefei 230026, China.}
\affiliation{CAS Center For Excellence in Quantum Information and Quantum Physics,
University of Science and Technology of China, Hefei, Anhui 230026,
China.}
\affiliation{Hefei National Laboratory, University of Science and Technology of China, Hefei 230088, China}

\author{Chang-Ling Zou}
\email{clzou321@ustc.edu.cn}
\affiliation{Laboratory of Quantum Information, University of Science and Technology of China, Hefei 230026, China.}
\affiliation{CAS Center For Excellence in Quantum Information and Quantum Physics,
University of Science and Technology of China, Hefei, Anhui 230026,
China.}
\affiliation{Hefei National Laboratory, University of Science and Technology of China, Hefei 230088, China}

\date{\today}
\renewcommand{\figurename}{Fig.}
\begin{abstract}
Stabilization of carrier-envelop-offset frequency ($f_{\mathrm{ceo}}$) in a microcomb typically relies on an additional optical or microwave reference. Here we propose autonomous frequency locking, in which Kerr comb formation and quadratic frequency conversion cooperate within a single microcavity to establish a zero-offset comb from one continuous-wave pump. The internal second-order interaction supplies phase-sensitive feedback that pins $f_{\mathrm{ceo}}$ to zero while the repetition rate self-adjusts to an integer division of the pump frequency. A second-order Adler-type equation links the locking threshold to the $\chi^{(2)}$ coupling strength and modal photon numbers. Simulations support this prediction and demonstrate recovery of the locked phase after perturbation. To explore a minimal realization of this mechanism, we further analyze a three-mode $\chi^{(3)}$ optical parametric oscillator whose signal and idler fields are coupled through $\chi^{(2)}$ conversion. Together, these interactions lock the signal and idler fields to $2/3$ and $4/3$ of the pump frequency. This configuration enables fractional-harmonic synthesis and provides a zero-offset frequency grid for measuring $f_{\mathrm{ceo}}$ with only one-third-octave spectral coverage. Our results establish internal frequency conversion as a route to autonomous microcomb control.

\end{abstract}

\maketitle

\setcounter{figure}{0} 
\renewcommand{\thefigure}{\textbf{\arabic{figure}}}
\renewcommand{\figurename}{\textbf{Fig}}

\emph{Introduction}.-~Integrated photonic devices have brought nonlinear optics into a new regime~\cite{shu2022microcomb,liu2022emerging,Wang2024LT}, where enhanced nonlinear interactions and wavelength-scale dispersion engineering allow multiple nonlinear processes involving multiple modes to occur simultaneously in a single configuration~\cite{Song2024,zhang2023second,Song2025,Stokowski2024,Yang2026DOPO,Yang2026Type0TypeI}. The cooperation of different nonlinear processes has attracted research interest in the exploration of new physics, such as the interference of $\chi^{(2)}$ and $\chi^{(3)}$ processes~\cite{Li2018,Li2018a,Cui2022,comb_SHG,RN223,Lu2023}, the competition between Raman/Brillouin scattering and Kerr processes~\cite{yu2020raman,bai2021brillouin,zhang2023soliton}, and it underlies chip-scale optical frequency combs~\cite{Bruch2021,Li2024breaking,hu2022high,yu2022femtosecond,QFYang2024,Ji2025,Kudelin2024,Sekine2025,Gray2026} and quantum light sources~\cite{Guidry2022,kues2019quantum}. When optical parametric gain is involved, such interactions can drive the intracavity field into phase-transition-like states of frequency locking and self-organization~\cite{Marte1994,Ropp2018,kondratiev2023recent,roy2023non}. In a pure $\chi^{(3)}$ cavity, cascaded four-wave mixing can self-organize the intracavity field into phase-coherent pulse trains and frequency combs known as dissipative Kerr solitons (DKSs)~\cite{Kippenberg2018,moille2020dissipative}. Such self-organization establishes equidistant comb lines but does not independently fix both the repetition frequency ($f_{\mathrm{rep}}$) and carrier-envelope offset frequency ($f_{\mathrm{ceo}}$), whose fluctuations must still be controlled for precision applications ~\cite{Brasch2017,Newman2019,Niu2023,Jin2024PRL,Jin2025}.

A soliton comb possesses two degrees of freedom, $f_{\mathrm{rep}}$ and $f_{\mathrm{ceo}}$, of which a single-frequency pump fixes only the combination $f_p=f_{\mathrm{ceo}}+n_p f_{\mathrm{rep}}$, where $n_p$ is the absolute pump mode number, leaving one frequency degree of freedom unconstrained. Most strategies developed so far remove this freedom by supplying a second constraint from outside the comb. A modulation tone imposed on the pump can trap the soliton in time and, when periodic, entrain $f_{\mathrm{rep}}$~\cite{Jang2015,Erkintalo2022,Todd2023}. A second laser can instead serve as a dichromatic pump~\cite{Taheri2017,Taheri2022}, be injected into a comb sideband~\cite{Wildi2023}, or synchronize the soliton through Kerr-mediated or mutual injection~\cite{Moille2023,Moille2026,Yang2017CP}, thereby pinning a comb line to its frequency. A similar scheme uses an optical parametric oscillator as a dual-point reference for all-optical frequency division~\cite{Zhao2024}. These approaches provide effective comb stabilization, and share a common principle, where the comb is locked by imposing an external frequency constraint, so its frequency, particularly \(f_{\mathrm{ceo}}\), remains determined by and inherits the fluctuations of the supplied reference.

In this Letter, we propose a new mechanism for autonomous frequency locking (AFL) through cooperative $\chi^{(3)}$ and $\chi^{(2)}$ interactions driven by a single continuous-wave laser. We first apply AFL to a multimode cavity supporting both dissipative Kerr solitons (DKSs) and second-harmonic generation (SHG). Intracavity $\chi^{(2)}$ conversion imposes a harmonic relation $f_{2n}=2f_n$, drives the repetition rate toward $f_p/n_p$, and thereby locks $f_{\mathrm{ceo}}$ to zero. A soliton-momentum analysis yields an extended Adler-type equation linking the locking threshold to the $\chi^{(2)}$ coupling strength and modal photon numbers, while simulations support the predicted lower locking boundary and demonstrate phase recovery after perturbation. A minimal three-mode OPO--SHG model further illustrates this mechanism and offers a simpler configuration for initial experimental investigation, generating exact fractional harmonics at $2f_p/3$ and $4f_p/3$. AFL thus establishes a known zero offset without an auxiliary frequency reference, $f_{\mathrm{ceo}}$ detection, or electronic feedback. By replacing an externally imposed frequency constraint with one generated by nonlinear dynamics, our work establishes a distinct physical principle for autonomous frequency control in integrated photonics.

\begin{figure}[t!]
\centering
\includegraphics[width=\linewidth]{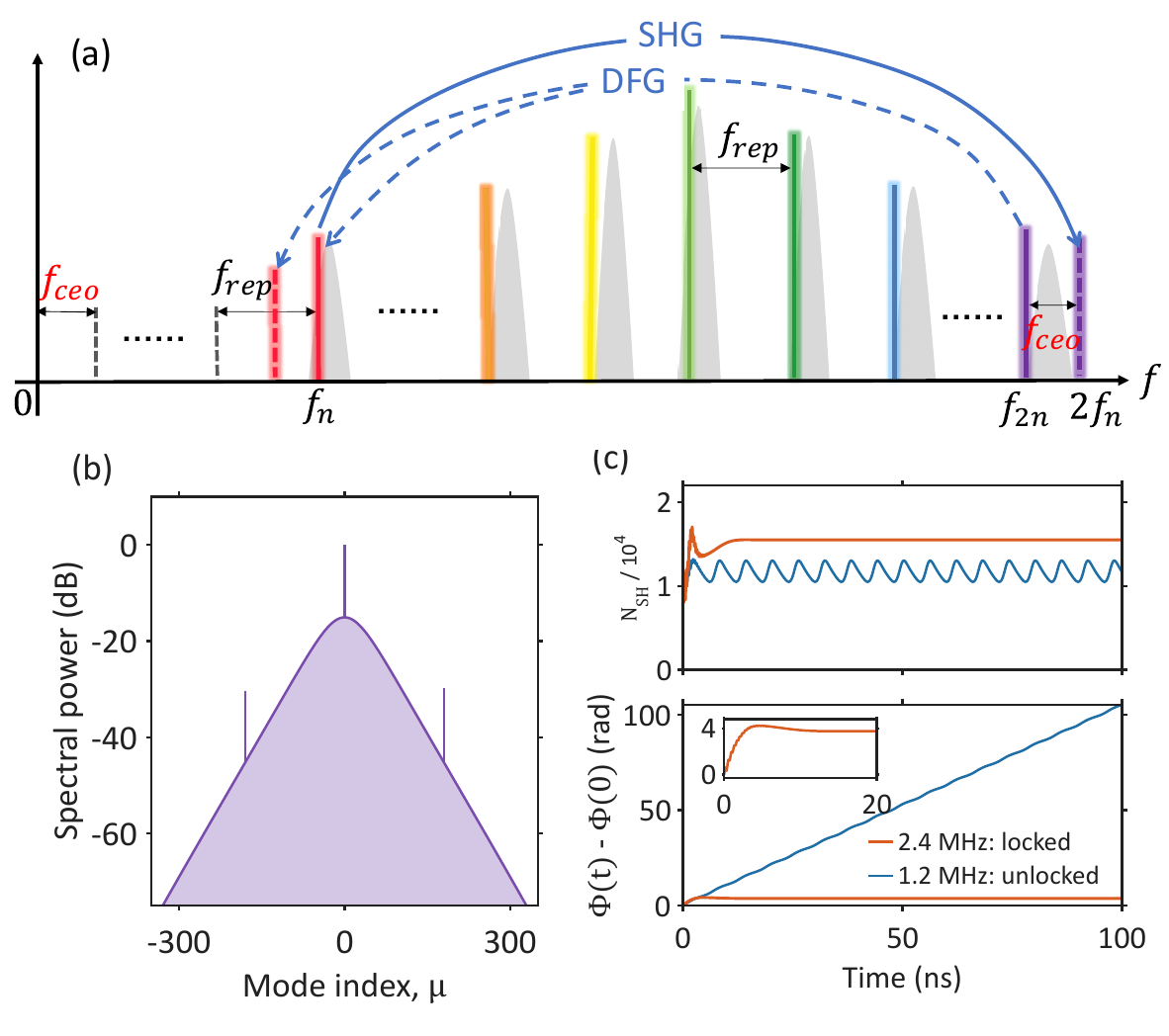}
\caption{Autonomous zero-$f_{\mathrm{ceo}}$ locking of a Kerr soliton.
(a) Intracavity $\chi^{(2)}$ couples the F and SH comb lines, and locking their frequencies to $2f_n=f_{2n}$ enforces $f_{\mathrm{ceo}}=0$. The comb teeth are schematic.
(b) Initial Kerr-soliton spectrum, normalized to the pump tooth, with the selected modes at $\mu_F=-180$ and
$\mu_{\mathrm{SH}}=180$ relative to the pump ($n_p=540$).
(c) SH photon number (upper) and unwrapped combination phase $\Phi=\arg\beta-2\arg\alpha$ (lower) after activating $\chi^{(2)}$. Both trials start from the same prepared soliton at $f_0=180\,\mathrm{MHz}$. For $g_2/(2\pi)=1.2\,\mathrm{MHz}$ (blue), the phase continues to drift; for $g_2/(2\pi)=2.4\,\mathrm{MHz}$ (orange), both quantities settle. The inset resolves the initial
phase relaxation of the locked state. The locked state also passes the whole-comb phase and perturbation-recovery tests described in SIV and Fig.~S1 of the Supplemental Material~\cite{SM}.}
\label{fig_comb_scheme}
\end{figure}

\emph{Zero-$f_{\mathrm{ceo}}$ soliton comb}.-~
Figure~\ref{fig_comb_scheme}(a) illustrates Kerr-comb generation through cascaded four-wave mixing (FWM). Kerr interactions can organize the intracavity field into a soliton with equally spaced comb lines. However, FWM alone does not fix the comb offset, and its energy-conservation relation $f_n+f_m=f_k+f_l$, with $n+m=k+l$, is satisfied for any $f_{\mathrm{ceo}}$ because the offset cancels upon substituting $f_j=f_{\mathrm{ceo}}+jf_{\mathrm{rep}}$. 
Environmental perturbations and pump-frequency drift can therefore change both $f_{\mathrm{rep}}$ and $f_{\mathrm{ceo}}$, with $\delta f_{\mathrm{ceo}}=\delta f_p-n_p\delta f_{\mathrm{rep}}$.

Here we introduce intracavity $\chi^{(2)}$ coupling between a fundamental (F) comb mode $a_{\mathrm{F}}$ and a second-harmonic (SH) mode $a_{\mathrm{SH}}$ . SHG generates a field at $2f_n$ [purple dashed line in Fig.~\ref{fig_comb_scheme}(a)], while difference-frequency generation (DFG) produces a field at $f_{2n}-f_n$ [red dashed line]. For nonzero $f_{\mathrm{ceo}}$, these generated fields differ from the corresponding comb lines by $|f_{\mathrm{ceo}}|$. In particular, $2f_n-f_{2n}=f_{\mathrm{ceo}}$. The participating cavity modes can therefore contain distinct frequency components, which also interact with the rest of the comb through FWM. Within the locking range, the phase-sensitive conversion drives the repetition rate to adjust until the generated fields coincide with the corresponding comb lines, satisfying $f_{2n}=2f_n$. This internally enforced harmonic relation thus locks $f_{\mathrm{ceo}}$ to zero, with $f_{\mathrm{rep}}=f_p/n_p$.

Using parameters representative of LiNbO$_3$ microrings~\cite{Lu:20,SM},
we numerically investigate autonomous zero-$f_{\mathrm{ceo}}$ locking. We first prepare a Kerr soliton without the $\chi^{(2)}$ interaction and locally shift the fundamental (F) and second-harmonic (SH) mode resonances to enhance their photon numbers for stronger $\chi^{(2)}$ conversion [Fig.~\ref{fig_comb_scheme}(b)]. Such local enhancement of comb-line power can arise from avoided mode crossings~\cite{Xue2015} or dispersive-wave generation~\cite{Liu2021}. The initial free-running offset is calibrated to $f_0=f_p-n_p f_{\mathrm{rep}}^{(0)}=180\,\mathrm{MHz}$, where $f_{\mathrm{rep}}^{(0)}$ is the repetition rate before activating $\chi^{(2)}$.

Figure~\ref{fig_comb_scheme}(c) compares the subsequent evolution at two coupling strengths, starting from the same prepared soliton. We assess locking by tracking the second-harmonic photon number and the phase combination $\Phi=\arg\beta-2\arg\alpha$, where $\alpha $ and $\beta$ are the complex amplitudes of the fundamental and second-harmonic modes, respectively. For a coherent comb, the phase drift in the zero-$f_{\mathrm{ceo}}$ reference frame satisfies $\dot{\Phi}=2\pi f_{\mathrm{ceo}}$, so a stationary $\Phi$ indicates locking to $f_{\mathrm{ceo}}=0.$ For $g_2/(2\pi)=1.2\,\mathrm{MHz}$, the SH photon number oscillates and the combination phase $\Phi$ continues to drift, indicating that the comb remains unlocked. For $g_2/(2\pi)=2.4\,\mathrm{MHz}$, both the photon number and the combination phase settle to stationary values. The comb phase evolution confirms $f_{\mathrm{ceo}}=0$, and recovery after a phase perturbation demonstrates a restoring response [Supplemental Material, SIV.2 and Fig.~S1~\cite{SM}]. These results show that intracavity $\chi^{(2)}$ conversion can autonomously adjust the repetition rate to $f_{\mathrm{rep}}=f_p/n_p$, establishing zero-offset locking without an external optical reference.

To quantify the locking capabilities, we define the AFL locking boundary as the largest positive free-running offset $f_0$ that locks to zero with phase recovery after activating $\chi^{(2)}$. The locking mechanism connects AFL to established descriptions of optical synchronization, such as sideband injection locking and Kerr-induced synchronization, which have been described using soliton momentum balance and extended Adler equations~\cite{Wildi2023,Moille2023}. In AFL, the additional phase constraint is supplied instead by intracavity frequency conversion between two comb modes, while for external injection drive, the same derivation recovers the coupling strength of injection locking [Supplemental Material, SIII.6, ~\cite{SM}], so the two scenarios share the same soliton response mechanism and differ in the source of the phase-sensitive drive.

Define the spectral momentum $P=\sum_\mu \mu |a_\mu|^2$, where $\mu$ is relative to the pump and $|a_\mu|^2$ is the modal photon number. For an approximately preserved soliton profile, dispersion converts the spectral shift described by P into a repetition-rate correction,
\begin{equation}
    \delta\omega_{\mathrm{rep}}\simeq \frac{D_2P}{N_s},
\end{equation}
where $D_2$ is the quadratic-dispersion coefficient and $N_s$ is the photon number of the soliton after removing the continuous-wave background. Each SHG event changes P by $\mu_{SH}-2\mu_F=n_p$, while four-wave mixing conserves P. With quadratic coupling strength $g_2$ and uniform photon-number decay rate 2$\kappa$, the coupled-mode equations give the evolution for P,
\begin{equation}
    \dot P=-2\kappa P-2n_pg_2|\alpha|^2|\beta|\mathrm{sin}(\Phi).
\end{equation}
Setting $\dot P=0$ determines the stationary value of P sustained by $\chi^{(2)}$ conversion against cavity loss, and hence the corresponding repetition-rate correction. A detailed reduction is given in the Supplemental Material, SIII ~\cite{SM}.

For approximately stationary internal spectral phases, the repetition-rate correction gives 
\begin{equation}
    \dot \Phi \simeq2\pi f_0-\frac{n_pD_2P}{N_s}.
\end{equation}

Differentiating this relation and substituting the evolution equation for P, while treating the modal photon numbers and $N_s$ as approximately constant, yields a second-order Adler-type equation 
\begin{equation}
    \ddot{\varphi}+2\kappa\dot{\varphi}
    =2\kappa\left(2\pi f_0-K_0\sin\varphi\right),
    \label{afl_extended_adler}
\end{equation}
where $\varphi=\Phi-\pi$ for $D_2, g_2 >0$ and
\begin{equation}
    K_0=
    \frac{n_p^2D_2g_2}{\kappa N_s}
    |\alpha|^2|\beta|.
    \label{afl_momentum_coupling}
\end{equation}
The $\alpha,\beta$ and $N_s$ in Eq.~\eqref{afl_momentum_coupling} are evaluated in the free-running initial state, with photon number notation $N_{F}=|\alpha|^2 $ and $N_{SH}=|\beta|^2$.

\begin{figure}[t!]
\centering
\includegraphics[width=1.05\linewidth]{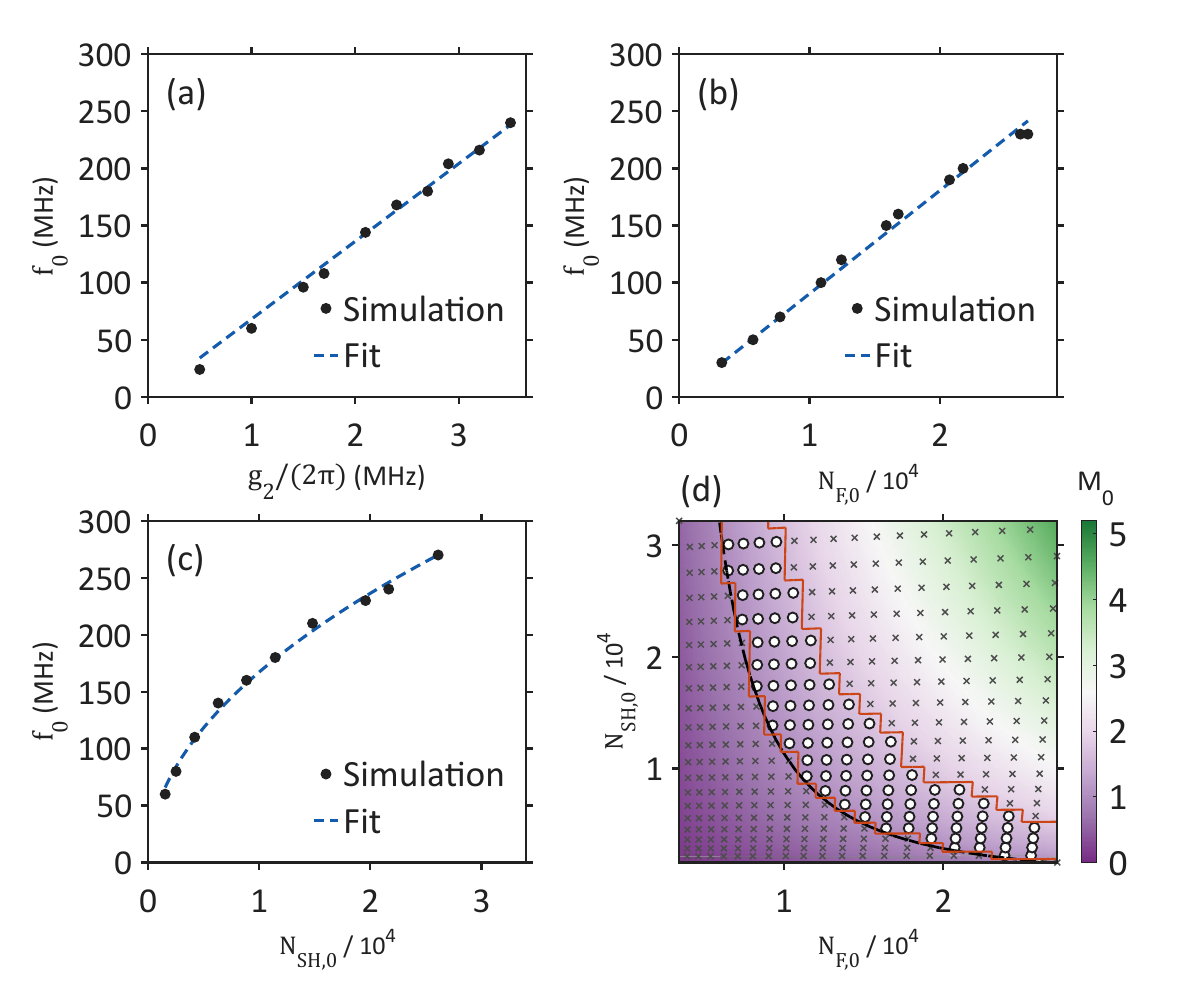}
\caption{
Numerical $f_{\mathrm{ceo}}$ locking boundaries and figure of merit for autonomous $f_{\mathrm{ceo}}$ locking.
(a) Free-running $f_{\mathrm{ceo}}$ locking boundary $f_0$ versus $g_2/(2\pi)$ at fixed initial photon numbers $N_F\simeq1.335\times10^4$ and $N_{\mathrm{SH}}\simeq5.238\times10^3$. $f_0/$MHz$= 68.04\cdot g_2/(2\pi)$.  
(b) Locking boundaries versus the initial fundamental photon numbers, at $g_2/(2\pi)=2.4\,\mathrm{MHz}$; the other mode remains within 1\% of $N_{\mathrm{SH}}=3000$. $f_0/$MHz$= 90.34 N_{F}/10^4$.      
(c) Locking boundaries versus the second-harmonic photon numbers, at $g_2/(2\pi)=2.4\,\mathrm{MHz}$; the other mode remains within 1\% of $N_F=10^4$. $f_0/$MHz$= 167.15 \sqrt{N_{SH}/10^4}$. 
(d) Initial-state figure of merit $M_0=K_0/(2\pi f_0)$, for 441 independently prepared states. White circles and black crosses denote 103 locked and 338 unlocked trials, respectively. The black contour indicates $M_0=1$, and the red contour estimates the locking boundary between adjacent sampled outcomes. } 
\label{fig_comb_lock}
\end{figure}

A stationary phase $\Phi$ requires $2\pi f_0=K_0\sin\varphi$, giving the maximum offset locking range $f_0=K_0/(2\pi)$. This range scales linearly with $g_2$ and $N_{F}$, and as the square root of $N_{SH}$, when the other parameters remain approximately constant. The numerical simulations verify these trends. At fixed initial soliton spectrum, the $f_{\mathrm{0}}$ locking boundary grows approximately linearly with $g_2$ [Fig.~\ref{fig_comb_lock}(a)]. At $g_2/(2\pi)=2.4\,\mathrm{MHz}$, varying the fundamental photon number while maintaining $N_{SH}\simeq3000$ gives an approximately linear trend [Fig.~\ref{fig_comb_lock}(b)], whereas varying the second-harmonic photon number at $N_{F}\simeq10^4$ gives an approximately square-root trend [Fig.~\ref{fig_comb_lock}(c)]. The dashed curves are descriptive fits constrained through the origin, with coefficients given in the caption. The fits agree well with the numerical results, supporting the theoretically predicted scaling of the locking range.

We therefore define the initial-state figure-of-merit
\begin{equation}
    \mathrm{FOM}\equiv M_0
    =\frac{K_0}{2\pi f_0}
    =\frac{n_p^2D_2g_2}
           {\kappa\,2\pi f_0}
      \frac{|\alpha|^2|\beta|}{N_s}.
    \label{combcondition}
\end{equation}
It compares the maximum $f_{\mathrm{ceo}}$ correction predicted by the momentum model with the free-running mismatch $f_0$. The reduced model admits a stable phase branch for $M_0>1$, with a marginal threshold at $M_0=1$. Figure~\ref{fig_comb_lock}(d) evaluates this quantity $M_0$ for 441 independently prepared states at $f_0=190\,\mathrm{MHz}$ and $g_2/(2\pi)=2.4\,\mathrm{MHz}$, using each state's own initial photon numbers. The numerical locking classification requires both stationary zero-$f_{\mathrm{ceo}}$ evolution and recovery after a phase perturbation, as detailed in the Supplemental Material, SIV ~\cite{SM}. The locking edge in the lower coupling regime with smaller photon numbers lies close to the $M_0=1$ contour, supporting the predicted onset of locking. However, $M_0>1$ is not sufficient for stationary locking. At larger photon numbers, perturbations from stronger $\chi^{(2)}$ conversion may extend beyond the F--SH mode pair through Kerr four-wave mixing and substantially reshape the soliton, invalidating the assumption of an approximately unchanged profile. The resulting oscillations can destabilize the stationary locked state [Supplemental Material, SIV ~\cite{SM}]. Nevertheless, our analysis identifies the minimum figure-of-merit required for locking, providing a quantitative criterion for assessing experimental feasibility.

\begin{figure}[t!]
\centering
\includegraphics[width=1\linewidth]{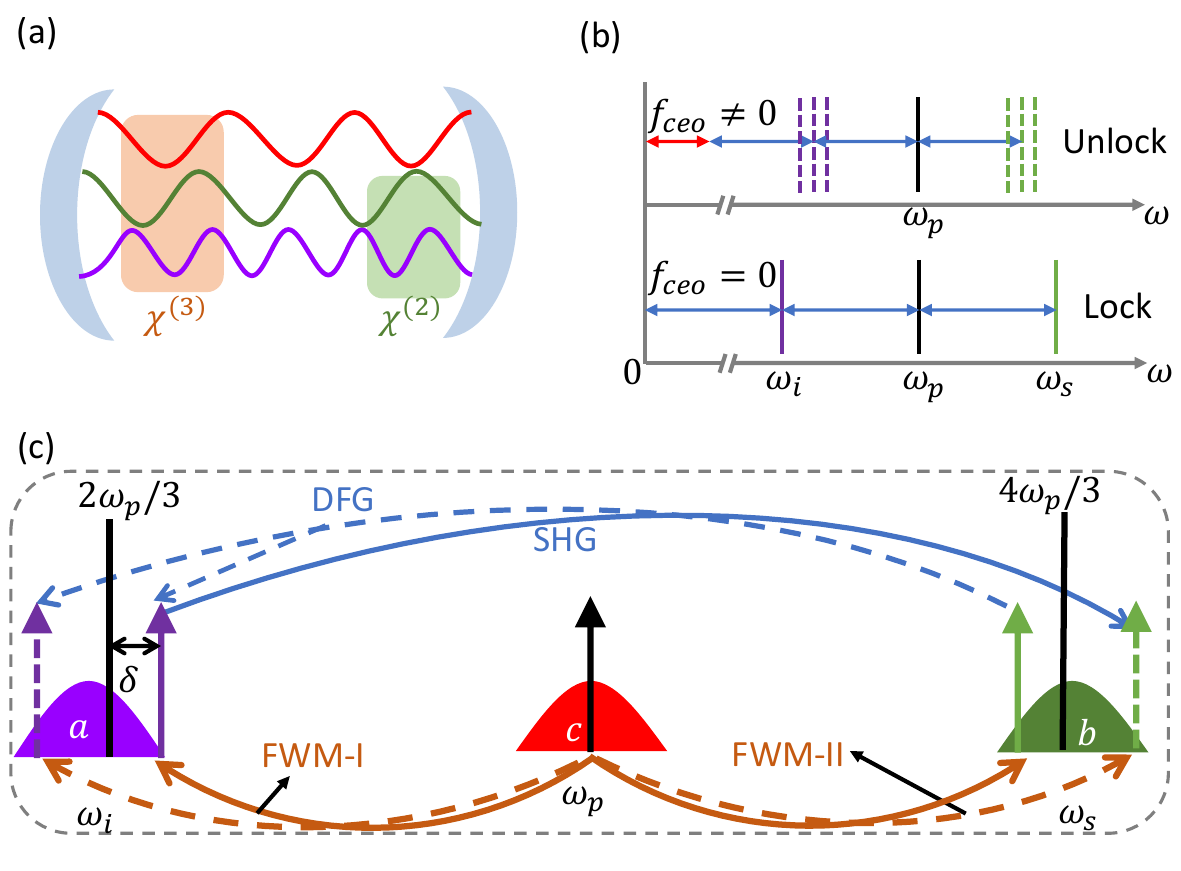}
\caption{Principle of the zero-offset frequency reference. (a) Illustration of a resonator for cascaded FWM ($\chi^{(3)}$) and SHG ($\chi^{(2)}$) processes. (b) Frequency tones generated by degenerate $\chi^{(3)}$ OPO without (upper) and with (lower) SHG. (c) Closed cascaded nonlinear optical network for an autonomously locked zero-offset reference. 
$\omega_{p}$: frequency of the pump laser, $\omega_{i,s}$: frequency of the OPO laser, FWM: four-wave mixing, SHG: second-harmonic generation, DFG: difference frequency generation.}
\label{fig_three_mode_scheme}
\end{figure}

\emph{Three-mode locking and fractional harmonics}.-~To give a clear physical insight into the mechanism of AFL, we 
consider a simple three-mode model that incorporates an $\chi^{(3)}$ degenerate OPO, i.e., the elementary process in Kerr comb formation and SHG, as shown in Fig.~\ref{fig_three_mode_scheme}(a). These processes, as individual components, have become the kernel elements of many integrated nonlinear photonic devices with exceptionally low power consumption~\cite{Lu:20,marty2021photonic,zhao2022ingap,JLin2019}. The degenerate OPO arises from the parametric interaction between three photonic modes, following the Hamiltonian
\begin{equation} \label{eq:FWM}
    H_{\mathrm{FWM}}	=	\sum_{j}\omega_{j}j^{\dagger}j+g_{3}\left(abc^{\dagger2}+a^{\dagger}b^{\dagger}c^{2}\right),
\end{equation}
where $j$ $\in\{a,b,c\}$ ($j^{\dagger}$) represents the annihilation (creation) operator of the mode, $g_{3}$ denotes the Kerr nonlinear coupling rate of OPO and $\omega_{j}$ corresponds to the mode frequency.

In the absence of $\chi^{(2)}$ process, the OPO is driven by an input laser $\omega_{p}$ in mode $c$. When operating above the OPO threshold, new lasing frequencies $\omega_{i,s}$ are initiated to form a three-line comb, satisfying $\omega_{i}+\omega_{s}=2\omega_{p}$. 
By choosing $\omega_{a,b}$ to be $\frac{2}{3}\omega_{p}$ and $\frac{4}{3}\omega_{p}$, it is anticipated that the pump, signal and idler lasers will form a three-line comb with zero frequency offset in an ideal scenario. 
However, due to environmental noises and pump frequency uncertainties, the detuning of lasing frequency $\omega_i$ with respect to the rotating frame $\frac{2}{3}\omega_{p}$~\cite{SM}
\begin{equation} \label{eq:OPO laser}
    \delta =\omega_{i}-\frac{2}{3}\omega_{p}	=	\frac{\Delta_{a}\kappa_{b}-\Delta_{b}\kappa_{a}}{\kappa_{a}+\kappa_{b}},
\end{equation}
fluctuates, with $\Delta_{a}=\omega_{a}-\frac{2}{3}\omega_{p}$, $\Delta_{b}=\omega_{b}-\frac{4}{3}\omega_{p}$ and $\kappa_a(b)$ is the dissipation rate of mode $a(b)$.

Similar to AFL in Kerr comb generation, this fluctuation can be eliminated by introducing SHG
\begin{equation} \label{eq:SHG}
    H_{\mathrm{SHG}}	=	g_{2}\left(a^{\dagger2}b+a^{2}b^{\dagger}\right),
\end{equation}
between modes $a$ and $b$ to establish a closed loop, where $g_{2}$ represents the nonlinear coupling rate of SHG. It imposes an additional constraint $\omega_{s}=2\omega_{i}$ on $\omega_{s,i}$. Together with $\omega_{i}+\omega_{s}=2\omega_{p}$, the OPO lasing frequencies are uniquely determined by $\omega_{p}$ if the system can achieve a steady state.

As shown in Fig.\,\ref{fig_three_mode_scheme}(c), when the OPO lasing frequency (purple solid arrow) is not exactly equal to $2\omega_{p}/3$ (black bar), the $\chi^{(2)}$ interaction [Eq.\,(\ref{eq:SHG})] generates an additional pair of sidebands (dashed green, purple arrows) located symmetrically around the pump. As a result, the beating of these lasing frequencies in each mode leads to periodic oscillation of the field intensity instead of a steady-state solution. However, the cooperation of $\chi^{(3)}$ interaction described by Eq.\,(\ref{eq:FWM})  further stimulates a process FWM-II, which competes with the original FWM-I and also contributes to the original OPO lines through the $\chi^{(2)}$ processes. The mutually reinforcing and competitive relationship means that a steady state solution is only permitted when $\delta=0$. By solving the dynamics of the cooperative nonlinear processes~ \cite{SM} we find the criteria for the existence of a steady state:
\begin{eqnarray}
9g_{2}^{4}\alpha^{4}-4\left(\Delta_{a}-\Delta_{b}\right)^{2}\kappa^{2}+4g_{2}^{2}\alpha^{2}\left(9\Delta_{a}\Delta_{b}+\kappa^{2}\right) & \geq & 0, \label{eq:condition}
\end{eqnarray}
where $\alpha^2$ is the photon number in mode $a$. A similar condition applies to mode $b$. For $|\Delta_{a,b}|\ll\kappa$ and $g_2|\alpha|\ll\kappa$, the criteria reduces to 
\begin{equation} \label{3modecondition}
    g_{2}^{2}\alpha^{2}	\geq	\left(\Delta_{a}-\Delta_{b}\right)^{2}.
\end{equation}
In this case, the AFL is realized as the lasing frequencies in modes $a$ and $b$ are autonomously locked to the fractional harmonics $2\omega_{p}/3$ and $4\omega_{p}/3$ of the pump laser, respectively. Consequently, the offset of the three-line comb is autonomously locked to zero, and remains robust against fluctuations in the pump or environment, in contrast to the unlocked OPO case [Fig.~\ref{fig_three_mode_scheme}(b)].

\begin{figure}[t!]
\centering
\includegraphics[width=\linewidth]{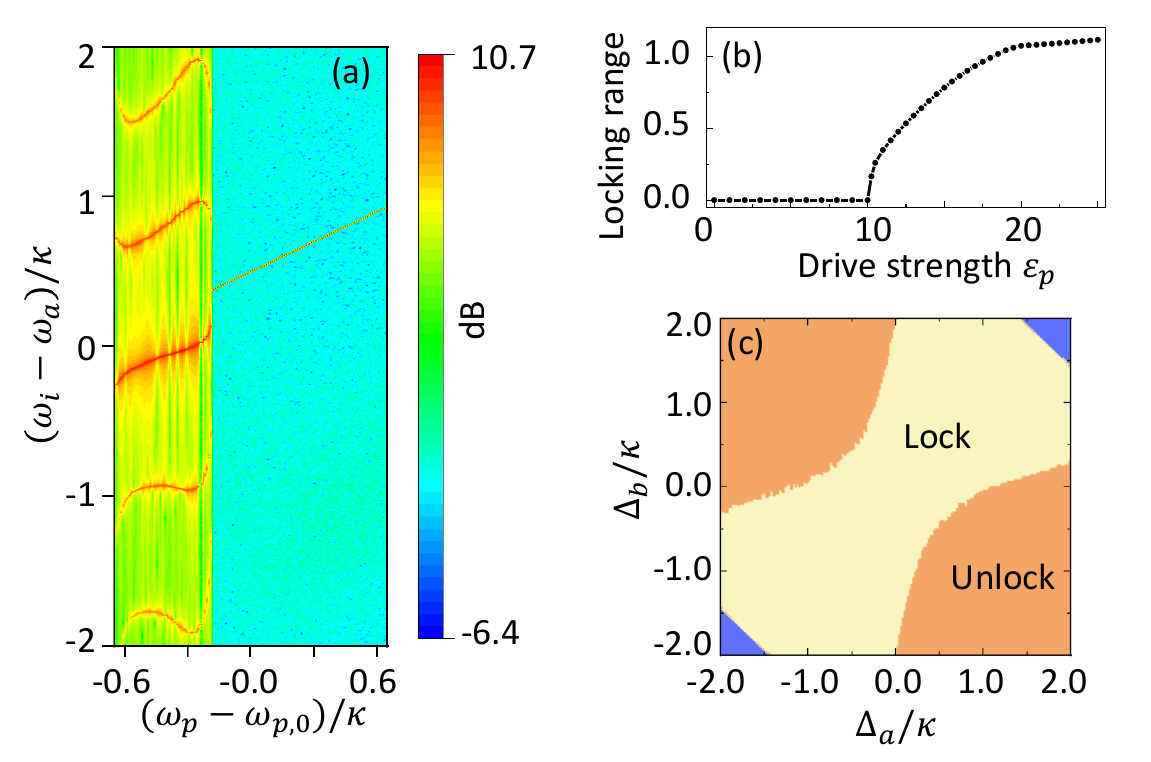}
\caption{ (a) Dynamics of the intracavity field spectrum of mode $a$ as the pump laser $\omega_{p}$ scans across the resonance. The OPO laser frequency $\omega_{i}$ has a single peak at the fractional harmonic frequency $2\omega_{p}/3$ after the system is locked. The initial detuning is $\Delta_{a}=-0.5\,\kappa$, $\Delta_{b}=0.3\,\kappa$, and $\kappa_{a}=\kappa_{b}=\kappa$. (b) Relationship between the drive strength $\varepsilon_{p}^{2}$ normalized to $10^{6}\kappa$ and the locking range normalized to $\kappa$. (c) Phase diagram of the system. The system operates below the OPO threshold inside the blue areas. In the calculation, the experimentally feasible parameters $g_{3}/\kappa=10^{-7}$ and $g_{2}/\kappa=10^{-3}$ are chosen.}
\label{fig_three_mode_lock}
\end{figure}

The AFL in three-line comb is verified numerically by analyzing the power spectral dynamics of the field as the pump laser frequency is scanned, as shown in Fig.~\ref{fig_three_mode_lock}(a). When the pump laser scans into the resonance of mode $c$, multiple lasing peaks emerge, as predicted by the FWM processes depicted in Fig.~\ref{fig_three_mode_scheme}(c). As anticipated, the $\chi^{(2)}$ interaction causes a transition from an unlocked multitone state to a locked single-tone state at a frequency equals exactly to $2/3\omega_{p}$, thus confirming the AFL mechanism. The parameter dependence of the existence of AFL is investigated, defining the locking range as the maximum detuning $\Delta_{b}$ that allows the locked state while fixing $\Delta_a=0$. It can be inferred from Eq.\,(\ref{3modecondition}) that a higher $g_{2}$ and a larger field amplitude $\alpha$ result in a larger locking range, where $\alpha$ can be enhanced by increasing the pump strength $\varepsilon_{p}$, as validated by  Fig.\,\ref{fig_three_mode_lock}(b) and is similar to Fig.\,\ref{fig_comb_lock}(a). Additionally, Fig. \ref{fig_three_mode_lock}(c) shows the phase diagram of the system for a fixed pump power $P_{in}/\hbar\omega_{p}=10^{7}\,\kappa_a$, with the stability being justified by the Routh-Hurwitz criterion \cite{DeJesus1987}. The AFL phase (yellow, Fig.\,\ref{fig_three_mode_lock}(c)) lies in the area where $\Delta_{a}$ and $\Delta_{b}$ have comparable values, which is consistent with Eq.~(\ref{3modecondition}).

The fixed frequency relation between the pump and OPO lasers indicates a reliable zero-offset frequency reference, which provides an efficient approach for determining the  $f_{\mathrm{ceo}}$ of another comb source, reminiscent of the traditional $f$-$2f$ self-referencing technique \cite{Hitachi2014,Brasch2017,Liu2021,chen2020chaos}, while avoiding the requirement of a demanding octave-spanning comb.
For the example in Fig.\,\ref{fig_three_mode_scheme}(c), the frequency-locked lasers $\omega_{p}$ and $\frac{4}{3}\omega_{p}$ can be used as references for  comb lines ($m_1$ and $m_2$), allowing the extraction of $f_{\text{ceo}}$ through the relation $f_{\text{ceo}}=(3m_2-4m_1)f_{\text{rep}}\mp4\Delta_1\pm3\Delta_2$, where $\Delta_1$ and $\Delta_2$ represent the respective beating frequencies. Our scheme based on AFL can accurately characterize a comb with only $\frac{1}{3}$-octave bandwidth, equivalent to the 3$f$-$4f$ self-referencing scheme.

\emph{Discussion and conclusion}.-Experimental realization requires tuning the initial free-running offset $f_0=f_p-n_p f_{\mathrm{rep}}^{(0)}$ into the locking range. Integrated-heater experiments have reported $f_{\mathrm{ceo}}$ tuning ranges exceeding $100\,\mathrm{GHz}$ in Si$_3$N$_4$ soliton microcombs~\cite{Moille2022Heaters} and approximately $21.6\,\mathrm{GHz}$ in octave-spanning devices~\cite{Weng2025}. These results establish post-fabrication tuning as a means of adjusting the initial offset, complementing dispersion design. The simplified three-mode OPO--SHG system offers a more direct route to an initial demonstration: it implements the same internal harmonic constraint without requiring an octave-spanning soliton. Its constituent processes, degenerate Kerr parametric oscillation~\cite{marty2021photonic} and efficient quasi-phase-matched SHG~\cite{Lu:20}, have already been demonstrated separately in optical microcavities. This configuration therefore provides a promising experimental starting point for AFL and its extension to multimode combs.

In conclusion, we have proposed and numerically validated AFL for Kerr soliton combs and a three-mode OPO--SHG system, in which internal $\chi^{(2)}$ frequency conversion supplies the phase-sensitive feedback that pins $f_{\mathrm{ceo}}$ to zero. A second-order Adler-type description yields a static locking condition, and numerical simulations support the predicted lower locking boundary and demonstrate recovery after a phase perturbation. The mechanism may also extend to platforms lacking an intrinsic second-order response. Externally applied static fields in silicon and photoinduced static fields in SiN provide routes to the required effective $\chi^{(2)}$ coupling~\cite{Timurdogan2017,Lu2021}. Time-dependent bias fields could further enable offset locking to a microwave reference. AFL thus fixes the offset to a known value without a separate offset-control loop. Looking ahead, referencing the remaining repetition rate to a known standard could enable an absolute comb frequency grid on a chip, offering a route to compact precision-metrology platforms with the frequency-referencing capabilities of conventional tabletop systems.

\smallskip{}

\begin{acknowledgments}
This work was funded by the National Key R\&D Program (Grant No.~2021YFF0603701), and the National Natural Science Foundation of China (Grants No. 12374361, No. 12293053, No. 62305214 and 92265210). It was also supported by the Fundamental Research Funds for the Central Universities, and the USTC Research Funds of the Double First-Class Initiative. The numerical calculations in this paper have been done on the supercomputing system in the Supercomputing Center of the University of Science and Technology of China. This work was partially carried out at the USTC Center for Micro and Nanoscale Research and Fabrication.
\end{acknowledgments}


\begin{thebibliography}{99}

\bibitem{shu2022microcomb}
H.~Shu, L.~Chang, Y.~Tao, B.~Shen, W.~Xie, M.~Jin, A.~Netherton, Z.~Tao, X.~Zhang, R.~Chen \emph{et~al.}, Microcomb-driven silicon photonic systems, Nature \textbf{605}, 457 (2022).

\bibitem{liu2022emerging}
J.~Liu, F.~Bo, L.~Chang, C.-H. Dong, X.~Ou, B.~Regan, X.~Shen, Q.~Song, B.~Yao, W.~Zhang \emph{et~al.}, Emerging material platforms for integrated microcavity photonics, Science China Physics, Mechanics \& Astronomy \textbf{65}, 104201 (2022).

\bibitem{Wang2024LT}
C.~Wang \emph{et~al.}, Lithium tantalate photonic integrated circuits for volume manufacturing, Nature \textbf{629}, 784 (2024).


\bibitem{Song2024}
Y.~Song, Y.~Hu, X.~Zhu, K.~Yang, and M.~Lon{\v{c}}ar, Octave-spanning Kerr soliton frequency combs in dispersion- and dissipation-engineered lithium niobate microresonators, Light: Science \& Applications \textbf{13}, 225 (2024).

\bibitem{zhang2023second}
L.~Zhang, X.~Wu, Z.~Hao, R.~Ma, F.~Gao, F.~Bo, G.~Zhang, and J.~Xu, Second-harmonic and cascaded third-harmonic generation in generalized quasiperiodic poled lithium niobate waveguides, Optics Letters \textbf{48}, 1906 (2023).

\bibitem{Song2025}
Y.~Song, Y.~Hu, M.~Lon{\v{c}}ar, and K.~Y. Yang, Hybrid Kerr-electro-optic frequency combs on thin-film lithium niobate, Light: Science \& Applications \textbf{14}, 270 (2025).

\bibitem{Stokowski2024}
H.~S. Stokowski \emph{et~al.}, Integrated frequency-modulated optical parametric oscillator, Nature \textbf{627}, 95 (2024).

\bibitem{Yang2026DOPO}
F.~Yang, J.~Xie, Y.~Zhou \textit{et al.}, Degeneracy-locked optical parametric oscillator, Nature Photonics \textbf{20}, 468--475 (2026).

\bibitem{Yang2026Type0TypeI}
F.~Yang and H.~X. Tang, Simultaneous Type-0 and Type-I optical parametric oscillation in submicron poled thin-film lithium niobate, Nanophotonics \textbf{15}, e70155 (2026).

\bibitem{Li2018}
M.~Li, C.-L. Zou, C.-H. Dong, X.-F. Ren, and D.-X. Dai, Enhancement of second-harmonic generation based on the cascaded second- and third-order nonlinear processes in a multimode optical microcavity, Phys. Rev. A \textbf{98}, 013854 (2018).

\bibitem{Li2018a}
M.~Li, C.-L. Zou, C.-H. Dong, and D.-X. Dai, Optimal third-harmonic generation in an optical microcavity with $\chi^{(2)}$ and $\chi^{(3)}$ nonlinearities, Opt. Express \textbf{26}, 27294 (2018).

\bibitem{Cui2022}
C.~Cui, L.~Zhang, and L.~Fan, In situ control of effective kerr nonlinearity with pockels integrated photonics, Nature Physics \textbf{18}, 497 (2022).

\bibitem{comb_SHG}
J.~Szabados, B.~Sturman, and I.~Breunig, Frequency comb generation threshold via second-harmonic excitation in $\chi^{(2)}$ optical microresonators, APL Photonics \textbf{5} (2020).

\bibitem{RN223}
A.~Villois and D.~V. Skryabin, Soliton and quasi-soliton frequency combs due to second harmonic generation in microresonators, Opt Express \textbf{27}, 7098 (2019).

\bibitem{Lu2023}
J.~Lu, D.~N. Puzyrev, V.~V. Pankratov, D.~V. Skryabin, F.~Yang, Z.~Gong, J.~B. Surya, and H.~X. Tang, Two-colour dissipative solitons and breathers in microresonator second-harmonic generation, Nature Communications \textbf{14}, 2798 (2023).


\bibitem{yu2020raman}
M.~Yu, Y.~Okawachi, R.~Cheng, C.~Wang, M.~Zhang, A.~L. Gaeta, and M.~Lon{\v{c}}ar, Raman lasing and soliton mode-locking in lithium niobate microresonators, Light: Science \& Applications \textbf{9}, 9 (2020).


\bibitem{bai2021brillouin}
Y.~Bai, M.~Zhang, Q.~Shi, S.~Ding, Y.~Qin, Z.~Xie, X.~Jiang, and M.~Xiao, Brillouin-kerr soliton frequency combs in an optical microresonator, Physical Review Letters \textbf{126}, 063901 (2021).

\bibitem{zhang2023soliton}
H.~Zhang, T.~Tan, H.-J. Chen, Y.~Yu, W.~Wang, B.~Chang, Y.~Liang, Y.~Guo, H.~Zhou, H.~Xia \emph{et~al.}, Soliton microcombs multiplexing using intracavity-stimulated brillouin lasers, Physical Review Letters \textbf{130}, 153802 (2023).

\bibitem{Bruch2021}
A.~W. Bruch, X.~Liu, Z.~Gong, J.~B. Surya, M.~Li, C.-L. Zou, and H.~X. Tang, Pockels soliton microcomb, Nature Photonics \textbf{15}, 21 (2021).

\bibitem{Li2024breaking}
M.~Li, X.-X. Xue, Y.-L. Zhang, X.-B. Xu, C.-H. Dong, G.-C. Guo, and C.-L. Zou, Breaking the efficiency limitations of dissipative kerr solitons using nonlinear couplers, Science China Physics, Mechanics \& Astronomy \textbf{67}, 234211 (2024).

\bibitem{hu2022high}
Y.~Hu, M.~Yu, B.~Buscaino, N.~Sinclair, D.~Zhu, R.~Cheng, A.~Shams-Ansari, L.~Shao, M.~Zhang, J.~M. Kahn \emph{et~al.}, High-efficiency and broadband on-chip electro-optic frequency comb generators, Nature Photonics \textbf{16}, 679 (2022).

\bibitem{yu2022femtosecond}
M.~Yu, D.~Barton~III, R.~Cheng, C.~Reimer, P.~Kharel, L.~He, L.~Shao, D.~Zhu, Y.~Hu, H.~R. Grant \emph{et~al.}, Integrated femtosecond pulse generator on thin-film lithium niobate, Nature \textbf{612}, 252 (2022).

\bibitem{QFYang2024}
Q.-F. Yang, Y.~Hu, V.~Torres-Company, and K.~Vahala, Efficient microresonator frequency combs, eLight \textbf{4}, 18 (2024).

\bibitem{Ji2025}
Q.-X. Ji \emph{et~al.}, Dispersive-wave-agile optical frequency division, Nature Photonics \textbf{19}, 624 (2025).

\bibitem{Kudelin2024}
I.~Kudelin \emph{et~al.}, Photonic chip-based low-noise microwave oscillator, Nature \textbf{627}, 534 (2024).

\bibitem{Sekine2025}
R.~Sekine, R.~M. Gray, L.~Ledezma, S.~Zhou, Q.~Guo, and A.~Marandi, Multi-octave frequency comb from an ultra-low-threshold nanophotonic parametric oscillator, Nature Photonics (2025).

\bibitem{Gray2026}
R.~M. Gray \emph{et~al.}, Topological soliton frequency comb in nanophotonic lithium niobate, Nature \textbf{652}, 76 (2026).

\bibitem{Guidry2022}
M.~A. Guidry, D.~M. Lukin, K.~Y. Yang, R.~Trivedi, and J.~Vu{\v{c}}kovi{\'c}, Quantum optics of soliton microcombs, Nature Photonics \textbf{16}, 52 (2022).

\bibitem{kues2019quantum}
M.~Kues, C.~Reimer, J.~M. Lukens, W.~J. Munro, A.~M. Weiner, D.~J. Moss, and R.~Morandotti, Quantum optical microcombs, Nature Photonics \textbf{13}, 170 (2019).

\bibitem{Marte1994}
M.~A. Marte, Competing nonlinearities, Physical Review A \textbf{49}, R3166 (1994).

\bibitem{Ropp2018}
C.~Ropp, N.~Bachelard, D.~Barth, Y.~Wang, and X.~Zhang, Dissipative self-organization in optical space, Nature Photonics \textbf{12}, 739 (2018).

\bibitem{kondratiev2023recent}
N.~M. Kondratiev, V.~E. Lobanov, A.~E. Shitikov, R.~R. Galiev, D.~A. Chermoshentsev, N.~Y. Dmitriev, A.~N. Danilin, E.~A. Lonshakov, K.~N. Min?kov, D.~M. Sokol \emph{et~al.}, Recent advances in laser self-injection locking to high-q microresonators, Frontiers of Physics \textbf{18}, 21305 (2023).

\bibitem{roy2023non}
A.~Roy, R.~Nehra, C.~Langrock, M.~Fejer, and A.~Marandi, Non-equilibrium spectral phase transitions in coupled nonlinear optical resonators, Nature Physics \textbf{19}, 427 (2023).

\bibitem{Kippenberg2018}
T.~J. Kippenberg, A.~L. Gaeta, M.~Lipson, and M.~L. Gorodetsky, Dissipative kerr solitons in optical microresonators, Science \textbf{361}, eaan8083 (2018).


\bibitem{moille2020dissipative}
G.~Moille, L.~Chang, W.~Xie, A.~Rao, X.~Lu, M.~Davanco, J.~E. Bowers, and K.~Srinivasan, Dissipative kerr solitons in a iii-v microresonator, Laser \& Photonics Reviews \textbf{14}, 2000022 (2020).


\bibitem{Brasch2017}
V.~Brasch, E.~Lucas, J.~D. Jost, M.~Geiselmann, and T.~J. Kippenberg, Self-referenced photonic chip soliton kerr frequency comb, Light: Science \& Applications \textbf{6}, e16202 (2017).

\bibitem{Newman2019}
Z.~L. Newman, V.~Maurice, T.~Drake, J.~R. Stone, T.~C. Briles, D.~T. Spencer, C.~Fredrick, Q.~Li, D.~Westly, B.~R. Ilic \emph{et~al.}, Architecture for the photonic integration of an optical atomic clock, Optica \textbf{6}, 680 (2019).

\bibitem{Niu2023}
R.~Niu, M.~Li, S.~Wan, Y.~R. Sun, S.-M. Hu, C.-L. Zou, G.-C. Guo, and C.-H. Dong, khz-precision wavemeter based on reconfigurable microsoliton, Nature Communications \textbf{14}, 169 (2023).

\bibitem{Jin2024PRL}
X.~Jin, Z.~Lv, L.~Yao, Q.~Gong, and Q.-F. Yang, Self-suppressed quantum-limited timing jitter and fundamental noise limit of soliton microcombs, Physical Review Letters \textbf{133}, 073801 (2024).

\bibitem{Jin2025}
X.~Jin \emph{et~al.}, Microresonator-referenced soliton microcombs with zeptosecond-level timing noise, Nature Photonics \textbf{19}, 630 (2025).

\bibitem{Jang2015}
J.~K. Jang, M.~Erkintalo, S.~Coen, and S.~G. Murdoch, Temporal tweezing of light through the trapping and manipulation of temporal cavity solitons, Nature Communications \textbf{6}, 7370 (2015).

\bibitem{Erkintalo2022}
M.~Erkintalo, S.~G. Murdoch, and S.~Coen, Phase and intensity control of dissipative Kerr cavity solitons, Journal of the Royal Society of New Zealand \textbf{52}, 149 (2022).

\bibitem{Todd2023}
C.~Todd, Z.~Li, S.~Coen, S.~G. Murdoch, G.-L. Oppo, and M.~Erkintalo, Dynamics of temporal Kerr cavity solitons in the presence of rapid parameter inhomogeneities: from bichromatic driving to third-order dispersion, Physical Review A \textbf{107}, 013506 (2023).

\bibitem{Taheri2017}
H.~Taheri, A.~B. Matsko, and L.~Maleki, Optical lattice trap for Kerr solitons, The European Physical Journal D \textbf{71}, 153 (2017).

\bibitem{Taheri2022}
H.~Taheri, A.~B. Matsko, L.~Maleki, and K.~Sacha, All-optical dissipative discrete time crystals, Nature Communications \textbf{13}, 848 (2022).

\bibitem{Wildi2023}
T.~Wildi, A.~Ulanov, N.~Englebert, T.~Voumard, and T.~Herr, Sideband injection locking in microresonator frequency combs, APL Photonics \textbf{8}, 120801 (2023).

\bibitem{Moille2023}
G.~Moille, J.~Stone, M.~Chojnacky, R.~Shrestha, U.~A. Javid, C.~Menyuk, and K.~Srinivasan, Kerr-induced synchronization of a cavity soliton to an optical reference, Nature \textbf{624}, 267 (2023).

\bibitem{Moille2026}
G.~Moille, P.~Shandilya, J.~Stone, R.~Beard, S.-C.~Ou, Z.~Li, M.~Harrington, K.~Liu, R.~Rockmore, C.~R.~Menyuk, D.~J.~Blumenthal, S.~P.~Krzyzewski, M.~Erkintalo, and K.~Srinivasan, Self-aligned optical microcomb emerging between octave-separated lasers, Nature \textbf{658}, 350--357 (2026).

\bibitem{Yang2017CP} Q.-F. Yang, X.~Yi, K.~Y. Yang, and K.~Vahala, Counter-propagating solitons in microresonators, Nature Photonics \textbf{11}, 560--564 (2017).

\bibitem{Zhao2024}
Y.~Zhao, J.~K. Jang, G.~J. Beals, K.~J. McNulty, X.~Ji, Y.~Okawachi, M.~Lipson, and A.~L. Gaeta, All-optical frequency division on-chip using a single laser, Nature \textbf{627}, 546 (2024).

\bibitem{SM}
Supplementary material for "autonomous frequency locking for zero-offset microcombs" .

\bibitem{Lu:20}
J.~Lu, M.~Li, C.-L. Zou, A.~A. Sayem, and H.~X. Tang, Toward 1\% single-photon anharmonicity with periodically poled lithium niobate microring resonators, Optica \textbf{7}, 1654 (2020).

\bibitem{Xue2015}
X.~Xue, Y.~Xuan, P.-H. Wang, Y.~Liu, D.~E. Leaird, M.~Qi, and A.~M. Weiner, {Normal-dispersion microcombs enabled by controllable mode interactions}, Laser \& Photonics Reviews \textbf{9}, L23 (2015).

\bibitem{Liu2021}
X.~Liu, Z.~Gong, A.~W. Bruch, J.~B. Surya, J.~Lu, and H.~X. Tang, Aluminum nitride nanophotonics for beyond-octave soliton microcomb generation and self-referencing, Nature communications \textbf{12}, 1 (2021).

\bibitem{marty2021photonic}
G.~Marty, S.~Combri{\'e}, F.~Raineri, and A.~De Rossi, Photonic crystal optical parametric oscillator, Nature Photonics \textbf{15}, 53--58 (2021).

\bibitem{zhao2022ingap}
M.~Zhao and K.~Fang, Ingap quantum nanophotonic integrated circuits with 1.5\% nonlinearity-to-loss ratio, Optica \textbf{9}, 258 (2022).

\bibitem{JLin2019}
J.~Lin, N.~Yao, Z.~Hao, J.~Zhang, W.~Mao, M.~Wang, W.~Chu, R.~Wu, Z.~Fang, L.~Qiao, W.~Fang, F.~Bo, and Y.~Cheng, Broadband quasi-phase-matched harmonic generation in an on-chip monocrystalline lithium niobate microdisk resonator, Physical Review Letters \textbf{122}, 173903 (2019).

\bibitem{DeJesus1987}
E.~X. DeJesus and C.~Kaufman, Routh-hurwitz criterion in the examination of eigenvalues of a system of nonlinear ordinary differential equations, Physical Review A \textbf{35}, 5288 (1987).

\bibitem{Hitachi2014}
K.~Hitachi, A.~Ishizawa, T.~Nishikawa, M.~Asobe, and T.~Sogawa, Carrier-envelope offset locking with a 2f-to-3f self-referencing interferometer using a dual-pitch ppln ridge waveguide, Optics Express \textbf{22}, 1629 (2014).

\bibitem{chen2020chaos}
H.-J. Chen, Q.-X. Ji, H.~Wang, Q.-F. Yang, Q.-T. Cao, Q.~Gong, X.~Yi, and Y.-F. Xiao, Chaos-assisted two-octave-spanning microcombs, Nature Communications \textbf{11}, 2336 (2020).

\bibitem{Moille2022Heaters}
G.~Moille, D.~Westly, E.~F. Perez, M.~Metzler, G.~Simelgor, and K.~Srinivasan, Integrated buried heaters for efficient spectral control of air-clad microresonator frequency combs, APL Photonics \textbf{7}, 126104 (2022).

\bibitem{Weng2025}
H.~Weng, H.~Tu, V.~Kumar, L.~Wang, A.~A. Afridi, Q.~Lu, D.~Skryabin, W.~Guo, and J.~F. Donegan, Thermally controlled dual-mode Si$_3$N$_4$ microresonators for generation of octave-spanning Kerr solitons, Physical Review Applied \textbf{23}, 034029 (2025).

\bibitem{Timurdogan2017}
E.~Timurdogan, C.~V. Poulton, M.~J. Byrd, and M.~R. Watts, Electric field-induced second-order nonlinear optical effects in silicon waveguides, Nature Photonics \textbf{11}, 200 (2017).

\bibitem{Lu2021}
X.~Lu, G.~Moille, A.~Rao, D.~A. Westly, and K.~Srinivasan, Efficient photoinduced second-harmonic generation in silicon nitride photonics, Nature Photonics \textbf{15}, 131 (2021).


\end{thebibliography}
\end{document}


\begin{center}
{\Large\bfseries Supplemental Material for\par}
\vspace{0.5em}
{\Large\bfseries ``Autonomous frequency locking for zero-offset microcombs''\par}
\vspace{1em}
{\normalsize Fengyan Yang,$^{1,2}$ ,Ming Li, $^{1,2,3}$ Juanjuan Lu,$^{4}$ Guang-Can Guo,$^{1,2,3}$ and Chang-Ling Zou$^{1,2,3,*}$\par}
\vspace{0.5em}
{\small
$^1$Laboratory of Quantum Information, University of Science and Technology of China, Hefei 230026, China\\
$^2$CAS Center for Excellence in Quantum Information and Quantum Physics, University of Science and Technology of China, Hefei 230026, China\\
$^3$Hefei National Laboratory, University of Science and Technology of China, Hefei 230088, China\\
$^4$School of Information Science and Technology, ShanghaiTech University, Shanghai 201210, China\\
$^*$Corresponding author: \href{mailto:clzou321@ustc.edu.cn}{clzou321@ustc.edu.cn}
}
\end{center}
\vspace{1em}

\section{Derivation of the OPO laser frequency}

We consider the OPO process based on degenerate four-wave mixing among
photonic modes $a,b,c$, which can be modeled by the Hamiltonian
\begin{eqnarray}
H & = & \omega_{a}a^{\dagger}a+\omega_{b}b^{\dagger}b+\omega_{c}c^{\dagger}c+\nonumber \\
 &  & g_{3}\left(abc^{\dagger2}+a^{\dagger}b^{\dagger}c^{2}\right)+i\varepsilon_{p}\left(c^{\dagger}e^{-i\omega_{p}t}-ce^{i\omega_{p}t}\right).
\end{eqnarray}
$\omega_{a,b,c}$ denotes the resonant frequency of the corresponding
mode and $\omega_{a,b}$ are close to $\frac{2}{3}\omega_{p}$ and
$\frac{4}{3}\omega_{p}$, respectively. $g_{3}$ is the nonlinear
coupling rate, $\varepsilon_{p}$ and $\omega_{p}$ are the driving
strength and frequency on mode $c$, respectively. When operating below the threshold, the photon amplitudes in mode $a,b$ are much weaker than that in the pump mode $c$ and the field in mode $c$
can be approximated to be $\gamma=\frac{\varepsilon_{p}}{-i(\omega_{c}-\omega_{p})-\kappa_{c}}$,
with $\kappa_{c}$ being the dissipation rate of the mode. Then, in
the symmetric rotating frame of $\frac{2}{3}\omega_{p}a^{\dagger}a+\frac{4}{3}\omega_{p}b^{\dagger}b$,
the Hamiltonian reduces to
\begin{eqnarray}
H & = & \Delta_{a}a^{\dagger}a+\Delta_{b}b^{\dagger}b+g_{\mathrm{eff}}\left(ab+a^{\dagger}b^{\dagger}\right),
\end{eqnarray}
where $g_{\mathrm{eff}}=g_{3}|\gamma^{2}|$. The dynamics of the operators
can be derived according to the Heisenberg-Langevin
\begin{eqnarray}
\frac{d}{dt}a & = & \left(-i\Delta_{a}-\kappa_{a}\right)a-ig_{\mathrm{eff}}b^{\dagger}+\sqrt{2\kappa_{a}}a_{in},\label{eq:anoise}\\
\frac{d}{dt}b & = & \left(-i\Delta_{b}-\kappa_{b}\right)b-ig_{\mathrm{eff}}a^{\dagger}+\sqrt{2\kappa_{b}}b_{in}.\label{eq:bnoise}
\end{eqnarray}
$a_{in}$ and $b_{in}$ are the input noise on the photonic modes.
Introducing the Fourier transform
\begin{eqnarray}
O\left(\omega\right) & = & \frac{1}{2\pi}\int O\left(t\right)e^{-i\omega t}dt,\\
O^{\dagger}\left(-\omega\right) & = & \frac{1}{2\pi}\int O^{\dagger}\left(t\right)e^{-i\omega t}dt,
\end{eqnarray}
the dynamical equations transform to
\begin{eqnarray}
0 & = & \left[-i\left(\Delta_{a}+\omega\right)-\kappa_{a}\right]a\left(\omega\right)-ig_{\mathrm{eff}}b^{\dagger}\left(-\omega\right)+\sqrt{2\kappa_{a}}a_{in}\left(\omega\right),\\
0 & = & \left[i\left(\Delta_{a}-\omega\right)-\kappa_{a}\right]a^{\dagger}\left(-\omega\right)+ig_{\mathrm{eff}}b\left(\omega\right)+\sqrt{2\kappa_{a}}a_{in}^{\dagger}\left(-\omega\right),\\
0 & = & \left[-i\left(\Delta_{b}+\omega\right)-\kappa_{b}\right]b\left(\omega\right)-ig_{\mathrm{eff}}a^{\dagger}\left(-\omega\right)+\sqrt{2\kappa_{b}}b_{in}\left(\omega\right),\\
0 & = & \left[i\left(\Delta_{b}-\omega\right)-\kappa_{b}\right]b^{\dagger}\left(-\omega\right)+ig_{\mathrm{eff}}a\left(\omega\right)+\sqrt{2\kappa_{b}}b_{in}^{\dagger}\left(-\omega\right).
\end{eqnarray}
In a compact form 
\begin{eqnarray}
\left[\begin{array}{cccc}
\Gamma_{a}^{+} & 0 & 0 & -ig_{3}|\gamma^{2}|\\
0 & \Gamma_{a}^{-} & ig_{3}|\gamma^{2}| & 0\\
0 & -ig_{3}|\gamma^{2}| & \Gamma_{b}^{+} & 0\\
ig_{3}|\gamma^{2}| & 0 & 0 & \Gamma_{b}^{-}
\end{array}\right]\left[\begin{array}{c}
a\left(\omega\right)\\
a^{\dagger}\left(-\omega\right)\\
b\left(\omega\right)\\
b^{\dagger}\left(-\omega\right)
\end{array}\right]+\left[\begin{array}{c}
\sqrt{2\kappa_{a}}a_{in}\left(\omega\right)\\
\sqrt{2\kappa_{a}}a_{in}^{\dagger}\left(-\omega\right)\\
\sqrt{2\kappa_{b}}b_{in}\left(\omega\right)\\
\sqrt{2\kappa_{b}}b_{in}^{\dagger}\left(-\omega\right)
\end{array}\right] & = & 0,
\end{eqnarray}
where $\Gamma_{a}^{\pm}=\mp i\left(\Delta_{a}\pm\omega\right)-\kappa_{a}$
and $\Gamma_{b}^{\pm}=\mp i\left(\Delta_{b}\pm\omega\right)-\kappa_{b}$.
The solution is 
\begin{eqnarray}
\left[\begin{array}{c}
a\left(\omega\right)\\
a^{\dagger}\left(-\omega\right)\\
b\left(\omega\right)\\
b^{\dagger}\left(-\omega\right)
\end{array}\right] & = & \mathrm{\left[\boldsymbol{M}\left(\omega\right)\right]}_{4\times4}\left[\begin{array}{c}
\sqrt{2\kappa_{a}}a_{in}\left(\omega\right)\\
\sqrt{2\kappa_{a}}a_{in}^{\dagger}\left(-\omega\right)\\
\sqrt{2\kappa_{b}}b_{in}\left(\omega\right)\\
\sqrt{2\kappa_{b}}b_{in}^{\dagger}\left(-\omega\right)
\end{array}\right].
\end{eqnarray}
The power spectral density is
\begin{eqnarray}
S_{a}\left(\omega\right) & = & \langle a^{\dagger}\left(\omega\right)a\left(\omega\right)\rangle\nonumber \\
 & = & \frac{\kappa_{a}}{\pi}\left|\boldsymbol{\mathrm{M}}_{12}\right|^{2}+\frac{\kappa_{b}}{\pi}\left|\boldsymbol{\mathrm{M}}_{14}\right|^{2},
\end{eqnarray}
where we have used the property of the quantum noise 
\begin{eqnarray}
\langle O_{in,i}^{\dagger}\left(\omega\right)O_{in,j}\left(\omega'\right)\rangle & = & 0,\\
\langle O_{in,i}\left(\omega\right)O_{in,j}^{\dagger}\left(\omega'\right)\rangle & = & \frac{1}{2\pi}\delta_{ij}\delta\left(\omega-\omega'\right),
\end{eqnarray}
for $O_{in,j}\in\{a_{in},b_{in}\}$. By solving the expression of
$\boldsymbol{M}\left(\omega\right)$, we find the power spectral diverges
when 
\begin{eqnarray}
\left[-i\left(\Delta_{a}+\omega\right)-\kappa_{a}\right]\left[i\left(\Delta_{b}-\omega\right)-\kappa_{b}\right] & = & g_{3}^{2}|\gamma^{4}|,
\end{eqnarray}
this equation gives the OPO threshold condition. Under the Fourier
convention of Eqs.~(S5) and (S6), its real root $\omega_*$ corresponds
to the lasing frequency $\omega_i=2\omega_p/3-\omega_*$. The lasing
frequency offset from $2\omega_p/3$ is therefore
\begin{eqnarray}
\delta_{s} & = & \frac{\Delta_{a}\kappa_{b}-\Delta_{b}\kappa_{a}}{\kappa_{a}+\kappa_{b}},
\end{eqnarray}
which depends on the mode detuning and dissipation.

\section{Condition of autonomous-locking}

For the cascaded OPO-SHG system with resonant drive on mode $c$,
the Hamiltonian in the rotating frame reads, 
\begin{eqnarray}
H & = & \Delta_{a}a^{\dagger}a+\Delta_{b}b^{\dagger}b+g_{3}\left(abc^{\dagger2}+a^{\dagger}b^{\dagger}c^{2}\right)+g_{2}\left(a^{2}b^{\dagger}+a^{\dagger2}b\right)+i\varepsilon_{p}\left(c^{\dagger}-c\right).
\end{eqnarray}
When operating above the threshold, the optical fields in mode $\{a,b,c\}$
can be treated as complex numbers $\{\alpha,\beta,\gamma\}$ for $g_{2,3}$
much smaller than the mode dissipation. The dynamics of the system
follows
\begin{eqnarray}
\frac{d}{dt}\alpha & = & \left(-i\Delta_{a}-\kappa_{a}\right)\alpha-ig_{3}\beta^{*}\gamma^{2}-2ig_{2}\alpha^{*}\beta\label{eq:a}\\
\frac{d}{dt}\beta & = & \left(-i\Delta_{b}-\kappa_{b}\right)\beta-ig_{3}\alpha^{*}\gamma^{2}-ig_{2}\alpha^{2}\label{eq:b}\\
\frac{d}{dt}\gamma & = & -\kappa_{c}\gamma-2ig_{3}\gamma^{*}\alpha\beta+\varepsilon_{p}.\label{eq:c}
\end{eqnarray}
Since the phase of $\gamma$ can be set by the drive field $\varepsilon_{p}$,
we take $\gamma$ to be real in the following derivation. The remaining phase
freedom is used to take $\alpha$ to be real.
From Eq.$\:$\ref{eq:b}, we get
\begin{eqnarray}
\gamma^{2} & = & \frac{ \left(-i\Delta_{b}-\kappa_{b}\right)\beta-ig_{2}\alpha^{2}}{ig_{3}\alpha^{*}} .
\end{eqnarray}
Substituting this expression into Eq.~\eqref{eq:a}, we obtain
\begin{eqnarray}
0 & = & \left(-i\Delta_{a}-\kappa_{a}\right)\alpha-ig_{3}\beta^{*}\frac{\left(-i\Delta_{b}-\kappa_{b}\right)\beta-ig_{2}\alpha^{2}}{ig_{3}\alpha^{*}}-2ig_{2}\alpha^{*}\beta\nonumber \\
 & = & \left(-i\Delta_{a}-\kappa_{a}\right)\alpha-\beta^{*}\frac{\left(-i\Delta_{b}-\kappa_{b}\right)\beta-ig_{2}\alpha^{2}}{\alpha^{*}}-2ig_{2}\alpha^{*}\beta
\end{eqnarray}
Then
\begin{eqnarray}
0 & = & \left(-i\Delta_{a}-\kappa_{a}\right)|\alpha|^{2}-\beta^{*}\left[\left(-i\Delta_{b}-\kappa_{b}\right)\beta-ig_{2}\alpha^{2}\right]-2ig_{2}\alpha^{*2}\beta\\
 & = & \left(-i\Delta_{a}-\kappa_{a}\right)|\alpha|^{2}-\left(-i\Delta_{b}-\kappa_{b}\right)|\beta|^{2}+ig_{2}\alpha^{2}\beta^{*}-2ig_{2}\alpha^{*2}\beta.
\end{eqnarray}
Writing $\beta=x+iy$, these equations can be separated into real and imaginary
parts,
\begin{eqnarray}
3g_{2}\alpha^{2}y+\kappa_{b}x^{2}+\kappa_{b}y^{2}-\kappa_a\alpha^{2} & = & 0\\
-g_{2}\alpha^{2}x+\Delta_{b}x^{2}+\Delta_{b}y^{2}-\Delta_{a}\alpha^{2} & = & 0.
\end{eqnarray}
The roots of these equations are
\begin{eqnarray}
x & = & \frac{1}{2g_{2}^{2}\left(9\Delta_{b}^{2}+\kappa_{b}^{2}\right)}\left[9g_{2}^{3}\Delta_{b}\alpha^{2}+2g_{2}\kappa_{b}\left(\Delta_{b}\kappa_{a}-\Delta_{a}\kappa_{b}\right)\right.\nonumber \\
 &  & \left.\mp3\sqrt{g_{2}^{2}\Delta_{b}^{2}\left(9g_{2}^{4}\alpha^{4}-4\left(\Delta_{b}\kappa_{a}-\Delta_{a}\kappa_{b}\right)^{2}+4g_{2}^{2}\alpha^{2}\left(9\Delta_{a}\Delta_{b}+\kappa_{a}\kappa_{b}\right)\right)}\right],\\
y & = & \pm\frac{1}{2g_{2}^{2}\Delta_{b}\left(9\Delta_{b}^{2}+\kappa_{b}^{2}\right)}\left[-3g_{2}^{3}\Delta_{b}\kappa_{b}\alpha^{2}+6g_{2}\Delta_{b}^{2}\left(\Delta_{b}\kappa_{a}-\Delta_{a}\kappa_{b}\right)\right.\nonumber \\
 &  & \left.+3\sqrt{g_{2}^{2}\Delta_{b}^{2}\left(9g_{2}^{4}\alpha^{4}-4\left(\Delta_{b}\kappa_{a}-\Delta_{a}\kappa_{b}\right)^{2}+4g_{2}^{2}\alpha^{2}\left(9\Delta_{a}\Delta_{b}+\kappa_{a}\kappa_{b}\right)\right)}\right].
\end{eqnarray}
For the simple case where $\kappa_a=\kappa_b$, it can be seen that 
\begin{eqnarray}
9g_{2}^{4}\alpha^{4}-4\left(\Delta_{a}-\Delta_{b}\right)^{2}\kappa^{2}+4g_{2}^{2}\alpha^{2}\left(9\Delta_{a}\Delta_{b}+\kappa^{2}\right) & \geq & 0
\end{eqnarray}
is required for real algebraic roots of $x$ and $y$. This is a
necessary existence condition within the preceding reduction; the pump
equation and dynamical stability must also be satisfied. For
$|\Delta_{a,b}|\ll\kappa$ and $g_2|\alpha|\ll\kappa$, the
leading-order condition reduces to

\begin{eqnarray}
g_{2}^{2}\alpha^{2} & \geq & \left(\Delta_{a}-\Delta_{b}\right)^{2}.
\end{eqnarray}

\section{Theory of autonomous zero-$f_{\mathrm{ceo}}$ locking}
\label{sec:afl_momentum}

We derive autonomous frequency locking (AFL) from the coupled-mode
equations by relating phase-sensitive photon conversion to the soliton
repetition rate. The exact spectral-momentum balance, combined with an
approximately unchanged soliton response, gives a second-order
Adler-type equation and a static locking threshold. Numerical evolution
then tests capture into the locked state and recovery after a phase
perturbation.

\subsection{Coupled-mode equations and phase convention}

Let $a_\mu$ be the intracavity amplitude of mode $\mu$, normalized so that
$|a_\mu|^2$ is its photon number. Mode indices are relative to the pump,
$\mu=0$, whose absolute mode number is $n_p$. We use the fixed zero-offset
reference grid
\begin{equation}
    \Omega_r=\frac{\omega_p}{n_p},\qquad
    \omega_\mu^{\mathrm{ref}}=\omega_p+\mu\Omega_r,
\end{equation}
where $\omega_p$ is the pump angular frequency. The fundamental (F) and
second-harmonic (SH) modes satisfy
\begin{equation}
    \mu_{\mathrm{SH}}-2\mu_F=n_p.
    \label{eq:si_mode_relation}
\end{equation}
Thus $2\omega_F^{\mathrm{ref}}=\omega_{\mathrm{SH}}^{\mathrm{ref}}$,
and their quadratic coupling has no explicit time dependence in this frame.
We write
\begin{equation}
    \alpha=a_{\mu_F},\qquad \beta=a_{\mu_{\mathrm{SH}}},\qquad
    N_F=|\alpha|^2,\qquad N_{\mathrm{SH}}=|\beta|^2.
\end{equation}

For uniform loss and a mode-independent Kerr coefficient, the classical
coupled-mode equations are
\begin{equation}
\begin{split}
    \dot a_\mu={}&-\left(\kappa+i\Delta_\mu\right)a_\mu
       -2ig_3 C_\mu-i\varepsilon_p\delta_{\mu0}+S_\mu,\\
    C_\mu={}&\sum_{k+l-m=\mu}a_k a_l a_m^*,\\
    S_\mu={}&-2ig_2\alpha^*\beta\,\delta_{\mu,\mu_F}
       -ig_2\alpha^2\,\delta_{\mu,\mu_{\mathrm{SH}}}.
\end{split}
\label{eq:si_cme}
\end{equation}
Here $g_2=g^{(2)}$ and $g_3=g^{(3)}$ are the quadratic and Kerr
coupling coefficients, and $\varepsilon_p$ is the pump-drive amplitude. The real detuning
$\Delta_\mu=\omega_\mu^{\mathrm{cav}}-\omega_\mu^{\mathrm{ref}}= \delta_p + (D_1-\Omega_r)\mu+ \sum_{n=2}^N\mu^nD_n/n!+\delta^{\mathrm{local}}_{\mu}$  ,
contains the pump detuning $\delta_p$, dispersion $D_n$, and local resonance corrections $\delta^{\mathrm{local}}_{\mu}, \{\mu=\mu_F,\mu_{SH}\}$ at F and SH modes. $\kappa$ is the
field-amplitude decay rate. Values quoted as $g_2/(2\pi)$, $D_2/(2\pi)$, and $\kappa/(2\pi)$ are expressed in Hz. We choose $g_2>0$ by a constant phase convention.

Define the F--SH combination phase
\begin{equation}
    \Phi=\arg\beta-2\arg\alpha.
    \label{eq:si_physical_phase}
\end{equation}
With optical fields proportional to $e^{-i\omega t}$, a coherent comb
whose lines satisfy $\omega_\mu=\omega_p+\mu\omega_{\mathrm{rep}}$ obeys
\begin{equation}
    \dot\Phi=2\omega_F-\omega_{\mathrm{SH}}
       =\omega_p-n_p\omega_{\mathrm{rep}}
       =2\pi f_{\mathrm{ceo}}(t).
    \label{eq:si_phase_ceo}
\end{equation}
We use the signed offset near zero. For a coherent comb, a stationary
combination phase therefore corresponds to $f_{\mathrm{ceo}}=0$.
During nonstationary evolution, the selected-pair phase can also change
through internal spectral dynamics, as discussed below.

Throughout the derivation, the constant
$\omega_{\mathrm{ceo}}\equiv \omega_p-n_p\omega_{\mathrm{rep}}^{(0)}=2\pi f_0,$ denotes the initial free-running mismatch, which is distinct from the instantaneous offset $f_{\mathrm{ceo}}(t)$. The superscript $(0)$ denotes
free-running operation with $g_2=0$.

\subsection{Exact phase-sensitive spectral-momentum balance}

The spectral momentum
\begin{equation}
    P=\sum_\mu\mu|a_\mu|^2
    \label{eq:si_momentum}
\end{equation}
describes the photon-number distribution relative to the pump \cite{Wildi2023}. Its normalization gives the photon-number-weighted mean mode index,
\begin{equation}
    \bar\mu=\frac{\sum_\mu\mu|a_\mu|^2}
                    {\sum_\mu|a_\mu|^2}
           =\frac{P}{N},\qquad
    N=\sum_\mu|a_\mu|^2.
    \label{eq:si_mean_mode}
\end{equation}
Thus $P>0$ places the weighted spectral average above the pump mode, and $P<0$ places it below. Differentiating $P$ gives
\begin{equation}
    \dot P=2\operatorname{Re}\sum_\mu\mu a_\mu^*\dot a_\mu.
\end{equation}

To evaluate the quadratic contribution, define $R$ as the rate of change of the SH photon number caused by $\chi^{(2)}$ conversion alone:
\begin{equation}
\begin{split}
    R&=2\operatorname{Re}\left[\beta^*(-ig_2\alpha^2)\right]\\
     &=-2g_2N_F\sqrt{N_{\mathrm{SH}}}\sin\Phi.
\end{split}
\label{eq:si_conversion_rate}
\end{equation}
The corresponding F photon-number conversion rate is $-2R$,
conserving $N_F+2N_{\mathrm{SH}}$ under the $\chi^{(2)}$
interaction according to the Manley--Rowe relations. Its contribution to $\dot P$ is therefore
\begin{equation}
    \mu_F(-2R)+\mu_{\mathrm{SH}}R=n_pR.
\end{equation}

Cavity loss also changes $P$. The field-decay term $-\kappa a_\mu$ gives a photon-number decay rate
\begin{equation}
    2\operatorname{Re}\left[
    a_\mu^*\left(-\kappa a_\mu\right)
    \right]
    =-2\kappa|a_\mu|^2.
\end{equation}
Weighting by the mode index and summing gives the loss contribution
\begin{equation}
    -2\kappa\sum_\mu\mu|a_\mu|^2=-2\kappa P.
\end{equation}
The detunings do not change modal photon numbers,
the pump acts at $\mu=0$, and Kerr mixing conserves the sum
of mode indices. These terms therefore contribute nothing
to $\dot P$. Adding cavity loss and quadratic conversion yields
\begin{equation}
\begin{split}
    \dot P
    &=-2\kappa P+n_pR\\
    &=-2\kappa P
      -2n_pg_2N_F\sqrt{N_{\mathrm{SH}}}\sin\Phi.
\end{split}
\label{eq:si_exact_balance}
\end{equation}
The photon numbers in this balance may depend on time.
Interference between the existing SH field $\beta$ and the generated increment $-ig_2\alpha^2dt$ makes the conversion phase dependent. Changing $\Phi$ therefore changes the direction and rate of photon transfer and the resulting spectral momentum.

\subsection{From spectral redistribution to repetition-rate correction}

To relate $P$ to the soliton response, we subtract the continuous-wave
background from the pump-mode amplitude and define
\begin{equation}
    N_s=\sum_{\mu\ne0}|a_\mu|^2+|a_0-a_{\mathrm{cw}}|^2,
    \label{eq:si_ns}
\end{equation}
where $a_{\mathrm{cw}}$ is the lower-intensity continuous-wave solution
at the same pump strength and detuning. This retains the soliton
contribution at the pump mode and excludes the uniform background from
$N_s$. The subtraction leaves $P$ unchanged because the background
occupies only $\mu=0$. Thus $P/N_s$ is the mean mode index of the
background-subtracted soliton spectrum.

For quadratic dispersion, the slope of the cavity resonance frequencies
is $d\omega_\mu^{\mathrm{cav}}/d\mu=D_1+D_2\mu$.
A change in the soliton's mean mode index therefore changes the
repetition rate associated with its spectral position. To quantify this
response, we relate the spectrum to the pulse motion, following the
approach used in sideband injection locking~\cite{Wildi2023}.

Write the intracavity envelope and background-subtracted photon
density as
\begin{equation}
    A(\theta,t)=\sum_\mu a_\mu(t)e^{i\mu\theta},
    \qquad
    \rho=|A-a_{\mathrm{cw}}|^2,
    \qquad
    N_s=\int\rho\,\frac{d\theta}{2\pi}.
\end{equation}
Here $a_{\mathrm{cw}}$ is uniform and time independent at fixed pump parameters. Since
$\partial_\theta^2e^{i\mu\theta}=-\mu^2e^{i\mu\theta}$, the modal dispersion term $-iD_2\mu^2a_\mu/2$ becomes $i(D_2/2)\partial_\theta^2A$. We therefore write
\begin{equation}
    \partial_t A
    =i\frac{D_2}{2}\partial_\theta^2A+\mathcal R,
\end{equation}
where $\mathcal R$ contains the remaining field-evolution terms.

Differentiating $\rho$ gives
\begin{equation}
\begin{aligned}
    \partial_t\rho
    &=2\operatorname{Re}
      \left[(A-a_{\mathrm{cw}})^*\partial_tA\right]\\
    &=-D_2\operatorname{Im}
      \left[(A-a_{\mathrm{cw}})^*\partial_\theta^2A\right]
      +\mathcal S,
\end{aligned}
\end{equation}
with $\mathcal S=2\operatorname{Re} [(A-a_{\mathrm{cw}})^*\mathcal R]$. Using
\begin{equation}
    \partial_\theta
    \left[(A-a_{\mathrm{cw}})^*\partial_\theta A\right]
    =|\partial_\theta A|^2
      +(A-a_{\mathrm{cw}})^*\partial_\theta^2A
\end{equation}
and noting that $|\partial_\theta A|^2$ is real, we obtain
\begin{equation}
    \partial_t\rho+\partial_\theta J=\mathcal S,
    \qquad
    J=D_2\operatorname{Im}
      \left[(A-a_{\mathrm{cw}})^*\partial_\theta A\right].
\end{equation}
Thus, $J$ describes the redistribution of the
background-subtracted photon density by quadratic dispersion;
$\mathcal S$ accounts for the remaining contributions.

To describe the position of a localized pulse, choose an unwrapped angular interval of length $2\pi$ whose endpoints lie away from the pulse, and define
\begin{equation}
    \theta_s=\frac{1}{N_s}
       \int\theta\rho\,\frac{d\theta}{2\pi}.
\end{equation}
Neglecting the boundary fluxes, integration of the density
equation gives
\begin{equation}
    \dot N_s=\int\mathcal S\,\frac{d\theta}{2\pi}.
\end{equation}
Differentiating the normalized pulse position and integrating
by parts then yields
\begin{equation}
\begin{aligned}
    \dot\theta_s
    &=\frac{1}{N_s}
      \int\theta\,\partial_t\rho\,\frac{d\theta}{2\pi}
      -\frac{\theta_s}{N_s}\dot N_s\\
    &=\frac{1}{N_s}\int J\,\frac{d\theta}{2\pi}
      +\frac{1}{N_s}\int(\theta-\theta_s)\mathcal S
        \,\frac{d\theta}{2\pi}.
\end{aligned}
\end{equation}
The term containing $-\theta_s\mathcal S$ comes from
differentiating $N_s$; no constant-photon-number approximation
has been made here.

Finally, Fourier orthogonality gives
\begin{equation}
\begin{aligned}
    \int(A-a_{\mathrm{cw}})^*\partial_\theta A
        \,\frac{d\theta}{2\pi}
    &=\int A^*\partial_\theta A\,\frac{d\theta}{2\pi}\\
    &=i\sum_\mu\mu|a_\mu|^2=iP.
\end{aligned}
\end{equation}
The background term vanishes because
$\int\partial_\theta A\,d\theta=0$ over one full period.
Hence $\int J\,d\theta/(2\pi)=D_2P$, giving
\begin{equation}
    \dot\theta_s=\frac{D_2P}{N_s}
      +\frac{1}{N_s}\int(\theta-\theta_s)\mathcal S
         \,\frac{d\theta}{2\pi}.
    \label{eq:si_position_moment}
\end{equation}
This separates the quadratic-dispersion contribution to
pulse motion from the remaining contributions.

For an approximately unchanged soliton, we hold $N_s$ and the second
term at their free-running values. Since
$\omega_{\mathrm{rep}}=\Omega_r+\dot\theta_s$, this gives
\begin{equation}
    \omega_{\mathrm{rep}}\simeq\omega_{\mathrm{rep}}^{(0)}
       +\frac{D_2P}{N_{s,0}}.
    \label{eq:si_mobility}
\end{equation}
Here $N_{s,0}$ is the soliton photon number before activating
$\chi^{(2)}$. The stationary free-running value of $P$ is zero by
Eq.~\eqref{eq:si_exact_balance}; any baseline pulse drift is included
in the measured $\omega_{\mathrm{rep}}^{(0)}$.
Equation~\eqref{eq:si_mobility} assumes that coupling produces little
change in the remaining pulse-motion contributions, including those
from higher-order dispersion, local resonance shifts, and radiation.

\subsection{Adler equation and phase stability}

For a coherent comb with approximately stationary internal spectral
phases, Eqs.~\eqref{eq:si_phase_ceo} and \eqref{eq:si_mobility} give
\begin{equation}
    \dot\Phi\simeq\omega_{\mathrm{ceo}}
       -\frac{n_pD_2}{N_{s,0}}P.
    \label{eq:si_phase_momentum}
\end{equation}
Setting $\dot\Phi=0$ determines the momentum needed to cancel the
initial mismatch,
\begin{equation}
    P_* = \frac{N_{s,0}\omega_{\mathrm{ceo}}}{n_pD_2}.
\end{equation}
Sustained locking additionally requires $\dot P=0$: phase-sensitive
conversion must replenish the momentum lost through cavity decay.
Together, $\dot\Phi=0$ and $\dot P=0$ determine the stationary
locked solution. We retain the time dependence to test whether small
phase perturbations decay back to this solution.

Differentiating Eq.~\eqref{eq:si_phase_momentum}, substituting
Eq.~\eqref{eq:si_exact_balance}, and approximating the photon numbers
by their initial free-running values gives
\begin{equation}
    \ddot\Phi+2\kappa\dot\Phi
       =2\kappa \omega_{\mathrm{ceo}}
       +\frac{2n_p^2D_2g_2}{N_{s,0}}
          N_{F,0}\sqrt{N_{\mathrm{SH},0}}\sin\Phi.
\end{equation}
For $D_2,g_2>0$, define
\begin{equation}
    \varphi=\Phi-\pi,\qquad
    K_0=\frac{n_p^2D_2g_2}{\kappa N_{s,0}}
         N_{F,0}\sqrt{N_{\mathrm{SH},0}}.
    \label{eq:si_k0}
\end{equation}
The phase equation becomes
\begin{equation}
    \ddot\varphi+2\kappa\dot\varphi
       =2\kappa\left(\omega_{\mathrm{ceo}}-K_0\sin\varphi\right).
    \label{eq:si_pendulum}
\end{equation}
This is a second-order Adler-type equation. The second derivative
accounts for the finite relaxation time of the spectral momentum.
The phase shift by $\pi$ gives the conventional sign of the sinusoidal
term and leaves $\dot\varphi=\dot\Phi$ unchanged. The coefficient
$K_0$ is the maximum steady-state correction to the angular offset
within this approximation.

A stationary phase satisfies
$\sin\varphi_*=\omega_{\mathrm{ceo}}/K_0$ and exists only for
$|\omega_{\mathrm{ceo}}|\le K_0$. To test its stability, write
$\varphi(t)=\varphi_*+\epsilon(t)$ with $|\epsilon|\ll1$.
Since $\varphi_*$ is constant, $\dot\varphi=\dot\epsilon$ and
$\ddot\varphi=\ddot\epsilon$. Expanding
\[
    \sin(\varphi_*+\epsilon)
      =\sin\varphi_*+\epsilon\cos\varphi_*+O(\epsilon^2)
\]
and substituting into Eq.~\eqref{eq:si_pendulum} gives
\[
    \ddot\epsilon+2\kappa\dot\epsilon
      =2\kappa\left[
        \omega_{\mathrm{ceo}}-K_0\sin\varphi_*
        -K_0\cos\varphi_*\,\epsilon\right]+O(\epsilon^2).
\]
The stationary condition cancels the first two terms. To linear order,
\begin{equation}
    \ddot\epsilon+2\kappa\dot\epsilon
      +2\kappa K_0\cos\varphi_*\,\epsilon=0.
    \label{eq:si_phase_perturbation}
\end{equation}
Taking $\epsilon\propto e^{\lambda t}$ yields
$\lambda^2+2\kappa\lambda+2\kappa K_0\cos\varphi_*=0$, with roots
\begin{equation}
    \lambda_\pm=
    -\kappa\pm
      \sqrt{\kappa^2-2\kappa K_0\cos\varphi_*}.
    \label{eq:si_eigenvalues}
\end{equation}
For $\cos\varphi_*>0$, both eigenvalues have negative real parts:
small phase perturbations decay and $f_{\mathrm{ceo}}$ returns to zero.
This branch is locally stable for $|\omega_{\mathrm{ceo}}|<K_0$.
At $|\omega_{\mathrm{ceo}}|=K_0$, the eigenvalues are $0$ and
$-2\kappa$; the linear restoring coefficient vanishes, and linearization
no longer guarantees recovery.

At $g_2=0$, the restoring coefficient vanishes. A translated
free-running soliton can remain at its new position even when $f_0=0$.
Recovery of the original phase therefore tests the restoring action
of the quadratic coupling.

\subsection{Figure of merit and limitations of the reduction}

For positive initial free-running offset $f_0$, we define
\begin{equation}
    M_0=\frac{K_0}{2\pi f_0}
       =\frac{n_p^2D_2g_2}{\kappa\,2\pi f_0}
         \frac{N_{F,0}\sqrt{N_{\mathrm{SH},0}}}{N_{s,0}}.
    \label{eq:si_fom}
\end{equation}
The numerator $K_0$ is the maximum angular-frequency correction,
and the denominator is the initial mismatch to be removed.
Equivalently, $M_0$ compares the maximum conversion-induced momentum
source with $2\kappa|P_*|$, the source needed to sustain the required
repetition-rate change. The reduced model requires $M_0>1$ for a
locally stable locked phase, with $M_0=1$ marking the static threshold.
This condition alone does not guarantee stationary capture, which
also requires stable soliton dynamics and a restoring phase response.
At $f_0=0$, the ratio is undefined and the restoring response is tested
directly.

At fixed initial spectrum, $K_0$ is linear in $g_2$. At approximately
constant $N_{s,0}$ and fixed other-mode photon number, it is linear in
$N_{F,0}$ and proportional to $\sqrt{N_{\mathrm{SH},0}}$.
The predicted symmetric static half-width is $K_0/(2\pi)$.
The capture range measured from a specified initial state can be
smaller than this static range.

The phase reduction assumes that the internal spectral phases remain approximately stationary, so that $\dot\Phi\simeq2\pi f_{\mathrm{ceo}}$.
The full coupled-mode equations also include changes in modal amplitudes and internal phases. These responses can modify the derived locking condition and are assessed through numerical
evolution.

\subsection{Relation to external injection locking}

The same connection between phase-sensitive driving, spectral
redistribution, and repetition-rate change underlies soliton sideband
injection locking~\cite{Wildi2023} and extended-Adler descriptions of
Kerr-induced synchronization~\cite{Moille2023}. For example, an external
drive $F_{\mathrm{inj}}e^{-i\delta_{\mathrm{inj}}t}$ on mode
$\mu_{\mathrm{inj}}$ contributes
\begin{equation}
    2\mu_{\mathrm{inj}}\operatorname{Re}
       \left[a_{\mu_{\mathrm{inj}}}^*F_{\mathrm{inj}}
       e^{-i\delta_{\mathrm{inj}}t}\right]
\end{equation}
to $\dot P$. Combining this source with the same repetition-rate
approximation gives a coupling magnitude
\begin{equation}
    K_{\mathrm{inj}}\simeq
    \frac{\mu_{\mathrm{inj}}^2|D_2|}{\kappa N_s}
       |F_{\mathrm{inj}}|\,|a_{\mu_{\mathrm{inj}}}|.
\end{equation}
External injection locks a comb line to an independent reference
laser. In AFL, the coupling is generated internally and acts on both
F and SH modes, with net spectral-momentum transfer weighted by
$n_p=\mu_{\mathrm{SH}}-2\mu_F$. It adjusts the repetition rate until
$2\omega_F=\omega_{\mathrm{SH}}$, giving $f_{\mathrm{ceo}}=0$.

\section{Numerical preparation and verification of locking}
\label{sec:comb_simulation}
\label{sec:afl_numerical_protocol}

\subsection{Soliton preparation and offset calibration}

We integrate Eq.~\eqref{eq:si_cme} with parameters representative of
an LN ring resonator. The targeted modes are $\mu_F=-180$ and
$\mu_{\mathrm{SH}}=180$, giving $n_p=540$.
Starting from an existing Kerr-soliton seed at $g_2=0$, we vary the
local F and SH resonance shifts and allow each state to relax.
The accepted states have stationary modal photon numbers and a
coherent spectrum; their common phase drift gives the free-running
repetition rate. Local resonance tuning changes the full soliton
solution, including the F and SH photon numbers. All photon numbers entering $M_0$ are
measured after this preparation and before activating $\chi^{(2)}$.
Figure~\ref{locking_diagnostics}(a,b) illustrates the preparation.

For each state, we measure its baseline free-running offset
$f_{\mathrm{base}}$ at $g_2=0$. At fixed pump frequency, a target offset $f_0$ is imposed by changing the cavity's first-order dispersion coefficient $D_1$, and thus the mode detuning $\Delta_\mu$. In the fixed zero-offset frame, this is implemented as
\begin{equation}
    \Delta_\mu(f_0)=\Delta_\mu^{\mathrm{base}}
       -\frac{2\pi(f_0-f_{\mathrm{base}})}{n_p}\,\mu.
    \label{eq:calibrated_scan}
\end{equation}
The pump detuning, higher-order dispersion, and additive local
resonance corrections remain fixed.

The calculations retain 701 modes and evaluate the Kerr convolution
with 2048-point zero padding to avoid modal-index aliasing. Time
integration uses fourth-order Runge--Kutta with a $1\,\mathrm{ps}$ step.
For the dense photon-number grid, preparation lasts $50\,\mathrm{ns}$,
followed by a $20\,\mathrm{ns}$ free-running calibration check and a
$100\,\mathrm{ns}$ evolution with $g_2$ enabled. Each capture trial
starts from its calibrated free-running state.

\subsection{Stationary capture and phase recovery}

\begin{figure}[t]
\centering
\includegraphics[width=\linewidth]{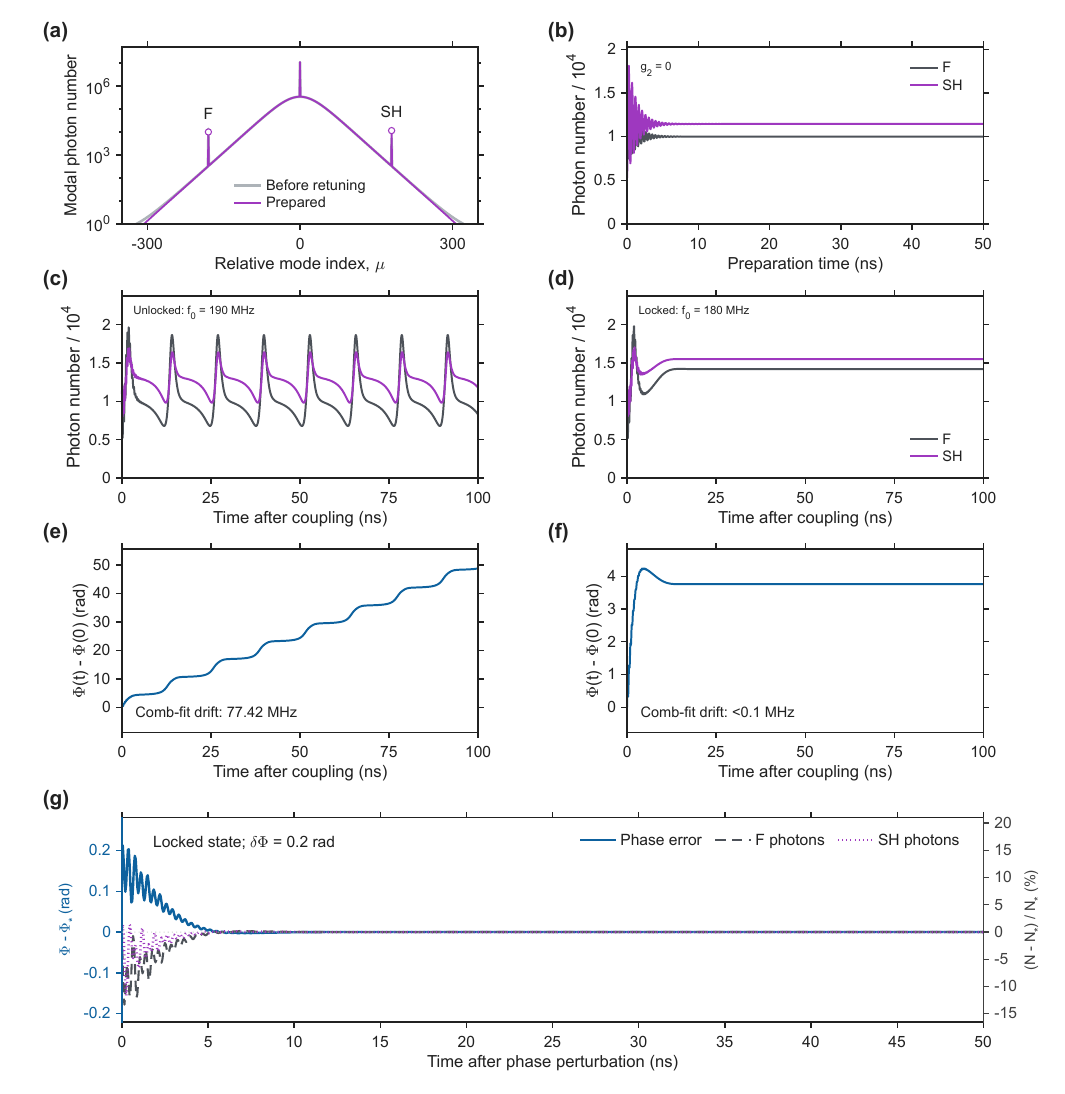}
\caption{Soliton preparation and verification of zero-$f_{\mathrm{ceo}}$
locking. (a) Photon-number spectra before and after local resonance
retuning of a Kerr-soliton seed with pre-existing local resonance
corrections. (b) Relaxation at $g_2=0$ to
$N_{F,0}\simeq1.001\times10^4$ and
$N_{\mathrm{SH},0}\simeq1.145\times10^4$.
(c--f) Photon numbers and unwrapped combination phase after enabling
$g_2/(2\pi)=2.4\,\mathrm{MHz}$ for the same prepared soliton:
(c,e) $f_0=190\,\mathrm{MHz}$, unlocked;
(d,f) $f_0=180\,\mathrm{MHz}$, locked.
(g) Recovery after the collective phase perturbation in
Eq.~\eqref{eq:collective_phase_perturbation}.
Phase error and relative photon-number changes are measured from
the pre-perturbation values.}
\label{locking_diagnostics}
\label{fig5}
\end{figure}

A candidate locked state must preserve a localized soliton and approach
a stationary field in the zero-offset frame. Its modal photon numbers
and F--SH combination phase must settle, and the full-comb phase
evolution must be consistent with
$f_{\mathrm{rep}}=f_p/n_p$, where $f_p=\omega_p/(2\pi)$.
These conditions require a coherent comb with no persistent pulse
drift. Persistent photon-number oscillations, including resolved
long-period oscillations, are classified as a failure of stationary
capture, even if the average phase drift is small. Representative
unlocked and locked trajectories are compared in
Fig.~\ref{locking_diagnostics}(c--f).

For candidates passing these tests, we perturb all modal phases by
\begin{equation}
    a_\mu\longrightarrow
    a_\mu\exp\left(i\mu\frac{\delta\Phi}{n_p}\right),\qquad
    \delta\Phi=0.2\,\mathrm{rad}.
    \label{eq:collective_phase_perturbation}
\end{equation}
This perturbation shifts the pulse position and increases the
combination phase $\Phi$ by $\delta\Phi$, while leaving the pump
phase and all modal photon numbers unchanged at the instant of
application. We then evolve the field with all parameters fixed
[Fig.~\ref{locking_diagnostics}(g)]. Recovery requires the field
to return to the same stationary state, with the combination phase
within $0.04\,\mathrm{rad}$ of its pre-perturbation value; phases
differing by an integer multiple of $2\pi$ are equivalent.
This recovery demonstrates a restoring response to the applied
phase perturbation.

\subsection{Comparison with the predicted locking threshold}

The one-dimensional scans test the predicted dependence on $g_2$, $N_{F,0}$, and $\sqrt{N_{\mathrm{SH},0}}$
[main-text Fig.~2(a--c)]. The coupling scan uses one common initial spectrum. The two photon-number scans use $g_2/(2\pi)=2.4\,\mathrm{MHz}$, with the other-mode photon number held within $1\%$ of its target: $N_{\mathrm{SH},0}\simeq3000$ or $N_{F,0}\simeq10^4$, respectively. 

For the two-dimensional map [main-text Fig.~2(d)], we vary both local resonance corrections and plot the resulting initial photon numbers. The 441 states use $g_2/(2\pi)=2.4\,\mathrm{MHz}$ and
$f_0=190\,\mathrm{MHz}$. The solver parameters are
$D_2/(2\pi)=0.7\,\mathrm{MHz}$ and
$\kappa/(2\pi)=95\,\mathrm{MHz}$.
Each point uses its own $N_{s,0}$ in Eq.~\eqref{eq:si_fom}, giving the $M_0=1$ contour without an adjustable threshold or scale factor. The low-coupling edge follows this contour closely. All sampled locked states have $M_0>1$, while many states with $M_0>1$ remain unlocked. Thus $M_0>1$ is necessary but not sufficient for stationary capture in the sampled data.

\begin{figure}[t]
\centering
\includegraphics[width=\linewidth]{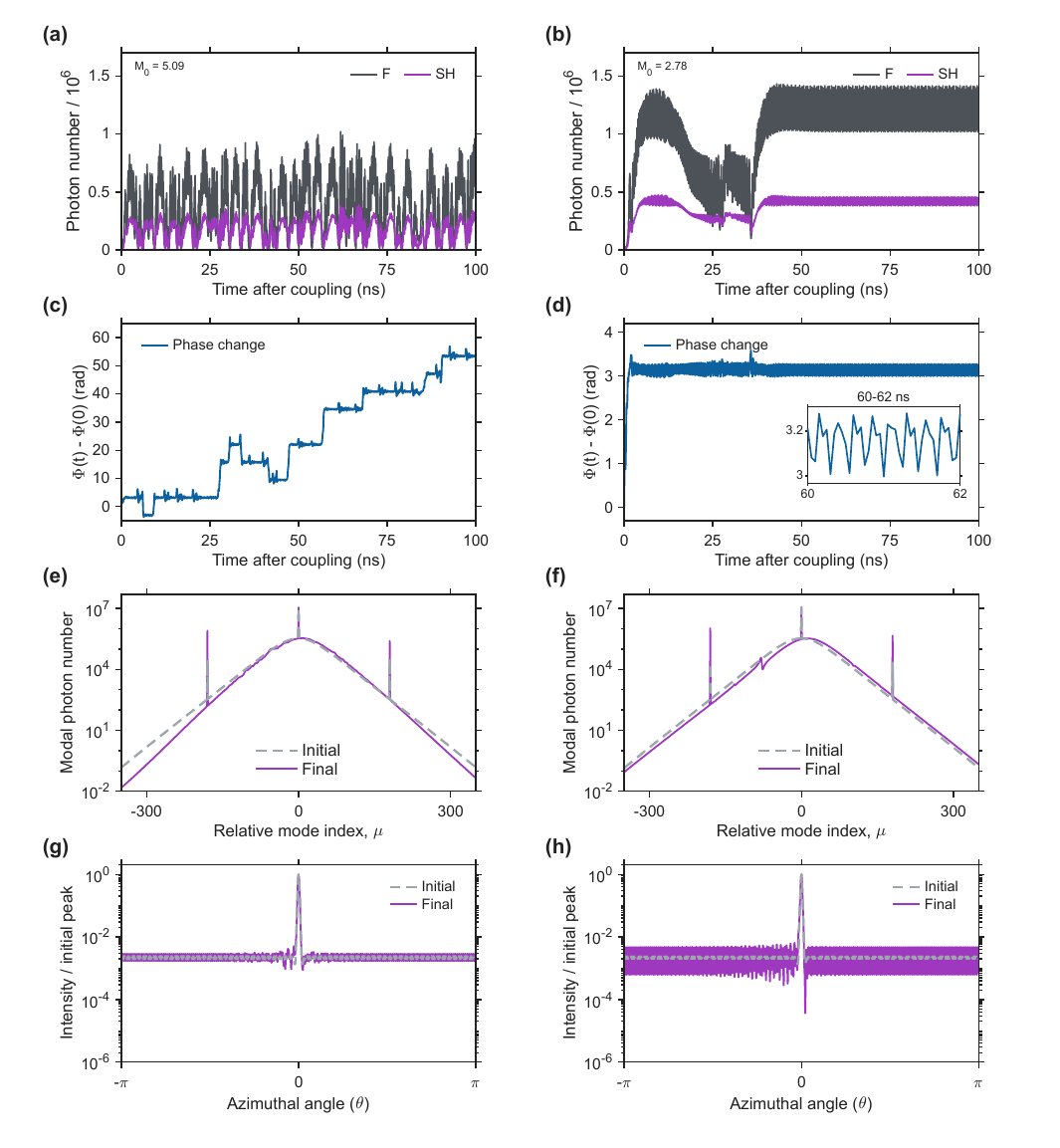}
\caption{Nonstationary states beyond the upper capture boundary at
$g_2/(2\pi)=2.4\,\mathrm{MHz}$ and $f_0=190\,\mathrm{MHz}$.
Initial photon numbers are
$(N_{F,0},N_{\mathrm{SH},0})\simeq(2.930,3.422)\times10^4$
in the left column and $(2.004,2.184)\times10^4$ in the right column.
(a,b) F- and SH-mode photon numbers.
(c,d) Unwrapped combination phase, showing phase slips and bounded
oscillations, respectively; the inset enlarges $60$--$62\,\mathrm{ns}$.
(e,f) Initial and final photon-number spectra.
(g,h) Initial and final pulse profiles over $\theta\in[-\pi,\pi]$,
aligned at their peaks and normalized to the initial peak intensity,
with logarithmic vertical axes.
Final fields are evaluated at $100\,\mathrm{ns}$.
Both cases retain localized pulses despite persistent amplitude
and phase dynamics.}
\label{fig:upper_boundary_dynamics}
\end{figure}

Beyond the upper capture boundary, persistent F- and SH-mode photon-number oscillations prevent stationary locking [Fig.~\ref{fig:upper_boundary_dynamics}(a,b)]. The two examples show phase slips and persistent phase oscillations within a finite range, respectively [Fig.~\ref{fig:upper_boundary_dynamics}(c,d)], while retaining broad comb spectra and localized pulses at the end
of the evolution [Fig.~\ref{fig:upper_boundary_dynamics}(e--h)]. The upper boundary therefore marks a limit of stationary capture without disappearance of the pulse. These dynamics fall outside the approximately constant photon numbers and stationary internal phases assumed in deriving $M_0$.

